\documentclass[prx,twocolumn,amsmath,amssymb,floatfix,footinbib,bibnotes,longbibliography]{revtex4-1}

\pdfoutput=1

\usepackage{soul}

\usepackage[colorlinks=true,citecolor=blue,linkcolor=magenta]{hyperref}
\usepackage{amsmath}
\usepackage{graphicx}
\usepackage{changepage}
\usepackage{fancyhdr}
\usepackage{amsthm, amssymb}
\usepackage{floatflt}
\usepackage{float}
\usepackage{array}
\usepackage[usenames,dvipsnames]{color}

\usepackage{rotating}
\usepackage{psfrag}
\usepackage{hyperref}
\usepackage{subfigure}
\usepackage{bm}% bold math
\usepackage{color}

\def \ra{{\rightarrow}}

\def \be{\begin{equation}}
\def \ee{\end{equation}}
\def \bea{\begin{eqnarray}}
\def \eea{\end{eqnarray}}
\def \nn{\nonumber}
\def \half{{1\over 2}}

\def \e{{\epsilon}}
\def \l{{\lambda}}
\def \L{{\Lambda}}
\def \a{{\alpha}}
\def \t{{\theta}}
\def \b{{\beta}}

\def \D{{\Delta}}
\def \d{{\delta}}
\def \w{{\omega}}
\def \s{{\sigma}}
\def \S{{\Sigma}}

\def \e{{\epsilon}}

\def \G{{\Gamma}}

\def \ra{{\rangle}}
\def \la{{\langle}}

\def \dag{{\dagger}}

\def \ba{\begin{align}}
\def \ea{\end{align}}

\def \half{{\frac{1}{2}}}

\def \mrm{\mathrm}

\def \mc{\mathcal}

\newcommand {\apgt} {\ {\raise-.5ex\hbox{$\buildrel>\over\sim$}}\ }
\newcommand {\aplt} {\ {\raise-.5ex\hbox{$\buildrel<\over\sim$}}\ }

\begin{document}

\title{Solvable model for a dynamical quantum phase transition from fast to slow scrambling }

\author{Sumilan Banerjee$^{1}$, Ehud Altman$^{1,2}$ \\
\small \em $^1$Department of Condensed Matter Physics, Weizmann Institute of Science, Rehovot 76100, Israel\\
$^{2}$ Department of Physics, University of California, Berkeley, CA 94720, USA}
\begin{abstract}
We propose an extension of the Sachdev-Ye-Kitaev (SYK) model that exhibits a quantum phase transition from the previously identified  non-Fermi liquid fixed point to a Fermi liquid like state, while still allowing an exact solution in a suitable large $N$ limit. The extended model involves coupling the interacting $N$-site SYK model to a new set of $pN$ peripheral sites with only quadratic hopping terms between them. The conformal fixed point of the SYK model remains a stable low energy phase below a critical ratio of peripheral sites  $p<p_c(n)$ that depends on the fermion filling $n$. The scrambling dynamics throughout the non-Fermi liquid phase is characterized by a universal Lyapunov exponent $\lambda_\mrm{L}\to 2\pi T$ in the low temperature limit, however the temperature scale marking the  crossover to the conformal regime vanishes continuously at the critical point $p_c$. The residual entropy at $T\to 0$, non zero in the NFL, also vanishes continuously at the critical point. For $p>p_c$ the quadratic sites effectively screen the SYK dynamics, leading to a quadratic fixed point in the low temperature and frequency limit. The interactions have a perturbative effect in this regime leading to scrambling with Lyapunov exponent $\lambda_\mrm{L}\propto T^2$. 
\end{abstract}
\maketitle

\section{Introduction}

Kitaev \cite{KitaevKITP} has recently given an intriguing new interpretation to a model of many fermions with all-to-all interactions,
introduced originally by Sachdev and Ye as a solvable example of a non-fermi liquid\cite{Sachdev1993}. The work of Sachdev and Ye  followed by Parcollet and Georges \cite{Parcollet1997} had investigated the saddle point solution, which is exact in the  thermodynamic (large $N$) limit and realizes a non-trivial conformal fixed point. The more recent work on this problem, initiated by Kitaev\cite{KitaevKITP,Polchinski2016,Maldacena2016,Bagrets2016}, used a simplified version henceforth called the SYK model, and focused on the dynamics leading to ergodicity, chaos and scrambling of quantum information. These studies uncovered a remarkable structure of the $1/N$ fluctuations in the SYK model and established a direct connection to quantum gravity with a black hole in $AdS_2$ \cite{Sachdev2010,KitaevKITP,Sachdev2015,Polchinski2016,Maldacena2016}. As in the case of a black hole, the scrambling in this system is characterized by a Lyapunov exponent $\lambda_\mrm{L}= 2\pi k_\mrm{B} T/\hbar$ that saturates the universal bound established in Ref. \cite{Maldacena2015}. In light of these results it is natural to ask if there is a broader classification of matter according to how it scrambles information. In particular one may ask if the SYK model can be tuned through a dynamical phase transition to a different state that does not scramble like a black hole.
\begin{center}
\begin{figure}
\includegraphics[width=0.5\textwidth]{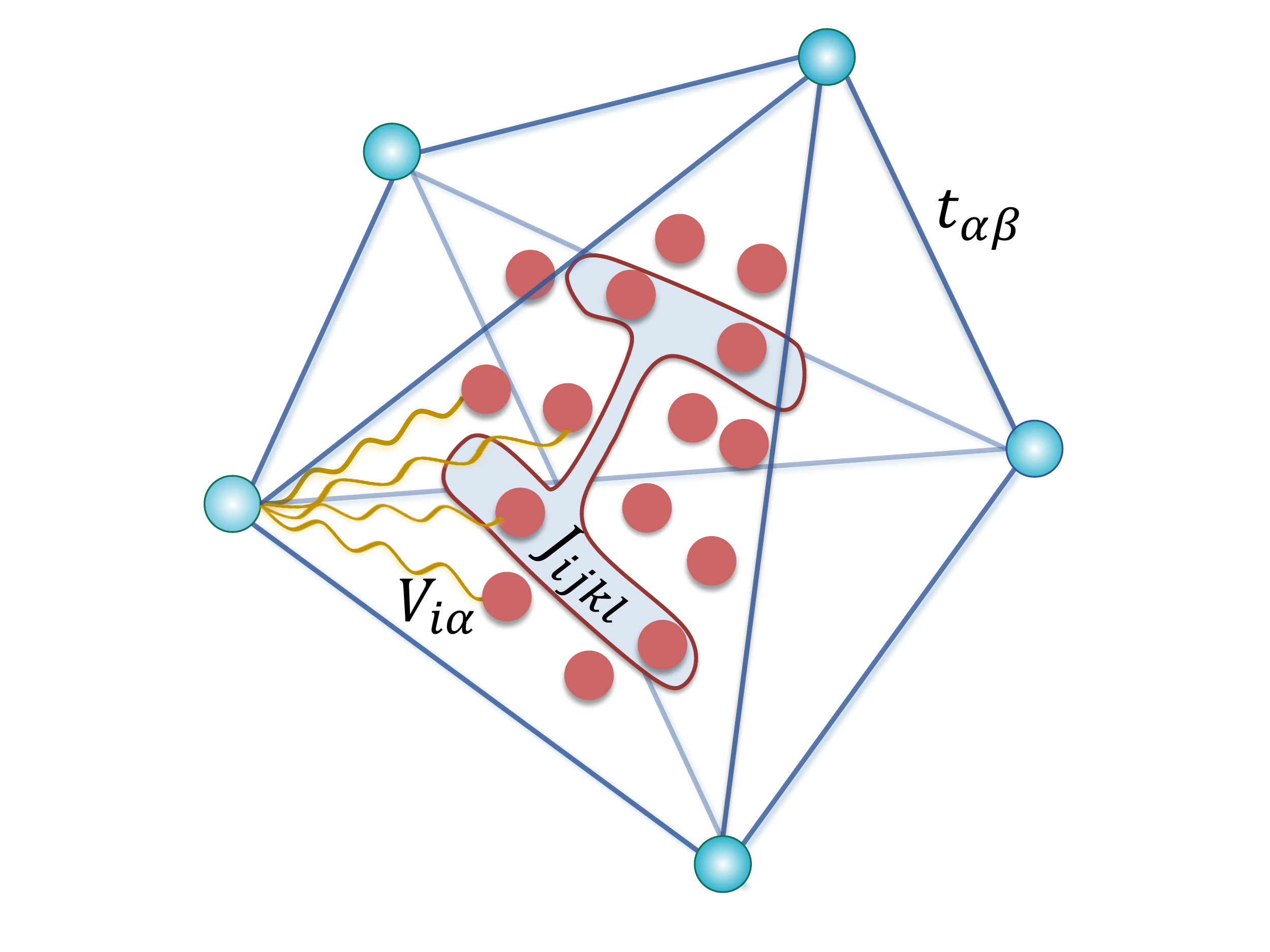}
\caption{The generalized SYK model. The SYK sites at the centre are coupled through random four fermion coupling $J_{ijkl}$. The sites at the periphery are connected to the SYK sites and to each other via random hoppings $V_{i\a}$ and $t_{\a\b}$, respectively.}
\label{fig.Model}
\end{figure}
\end{center}

\vspace{-2cm}
Here we begin to address these questions by generalizing the SYK model in a way that allows to drive a quantum phase transition between two low-energy fixed points. The two fixed points entail very different scrambling dynamics, which can be computed exactly in the large $N$ limit. Specifically we consider the model shown schematically in Fig.\ref{fig.Model}, with two species of fermions: 
\bea
\mc{H}=\mc{H}_c+\mc{H}_{\psi}+\mc{H}_{c\psi} \label{eq.Model}
\eea
where
\begin{subequations} \label{eq.Model_1}
\bea
\mc{H}_c&=&\frac{1}{(2N)^{3/2}}\sum_{ijkl}J_{ijkl}c_{i}^{\dagger}c_{j}^{\dagger}c_{k}c_{l}
-\mu\sum_{i}c_{i}^{\dagger}c_{i} \label{eq.SYK}
\\
\mc{H}_{\psi}&=&\frac{1}{M^{1/2}}\sum_{\alpha\beta}t_{\alpha\beta}\psi_{\alpha}^{\dagger}\psi_{\beta}
-\mu\sum_{\alpha}\psi_{\alpha}^{\dagger}\psi_{\alpha}\label{eq.Anderson}\\
\mc{H}_{c\psi}&=&\frac{1}{(NM)^{1/4}}\sum_{i\alpha}(V_{i\alpha}c_{i}^{\dagger}\psi_{\alpha}+V_{i\alpha}^{*}\psi_{\alpha}^{\dagger}c_{i})\label{eq.Coupling}
\eea 
\end{subequations}
The $c$ fermions, on sites $i=1,\dots,N$, are described by the SYK model [Eq.\eqref{eq.SYK}] with random four-fermion coupling $J_{ijkl}$ drawn from a Gaussian distribution with zero mean and variance $\overline{|J_{ijkl}|^2}=J^2$; $J_{ijkl}$ are properly antisymmetrized, i.e. $J_{ijkl}=-J_{jikl}=-J_{ijlk}$ and $J_{ijkl}=J_{klij}^*$. Here we have adopted a version \cite{Sachdev2015} of SYK model with complex fermions where one can tune the fermion density by changing the chemical potential $\mu$. The $\psi$ fermions reside on a separate set of ``peripheral" sites $\alpha=1,\dots,M$ connected with each other via hopping $t_{\a\b}$. Finally, there is a coupling $V_{i\a}$ between the two species of fermions. Both $t_{\a\b}$ and $V_{i\a}$ are complex Gaussian random variable with zero mean and variances $\overline{|t_{\a\b}|^2}=t^2$ and $\overline{|V_{i\a}|^2}=V^2$, respectively. The $N$ and $M$ dependent prefactors in equations \eqref{eq.Model_1} ensure proper thermodynamic limit for $N,M\to\infty$ with a fixed ratio $M/N=p$. 

Without the term in equation \eqref{eq.Coupling}, the model describes two decoupled systems. The SYK Hamiltonian $\mc{H}_c$ is solvable in the large-$N$ limit and has an emergent conformal symmetry at low energies \cite{KitaevKITP,Maldacena2016,Sachdev2015}. As mentioned above, the model gives rise to thermalization and many-body quantum chaos with Lyapunov exponent, $\lambda_\mrm{L}=2\pi T$ ($k_\mrm{B}=1,~\hbar=1$), that saturates the quantum limit, like a black hole in Einstein gravity \cite{KitaevKITP,Maldacena2015}. The quadratic model alone is also solvable and of course it does not exhibit scrambling. 

We show that the low energy dynamics in the coupled system is crucially determined by the ratio of peripheral sites to SYK sites $p=M/N$. If $p$ is smaller than a critical value $p_c$, then the dynamics is still controlled by a strong coupling SYK-like fixed point with the universal Lyapunov exponent $\lambda_\mrm{L}=2\pi T$. On the other hand, for $p>p_c$ the quadratic fermions effectively screen the SYK interactions leading to a free low-energy fixed point characterized by a non universal Lyapunov exponent $\lambda_\mrm{L}\propto T^2$. 
These two phases are separated by a continuous quantum phase transition at $p=p_c$.
Interestingly, the critical ratio $p_c$ depends only on the fermion density and not on the coupling strength $V^2$. The latter only sets the energy scale below which the two fermion species are effectively coupled.  

We should stress the difference between this model and another natural extension of the SYK model. It is tempting to consider a model with both quadratic and quartic couplings of fermions on the same $N$ sites. Since a model with only quadratic coupling between the sites gives rise to a different fixed point than the SYK model, one might naively expect a phase transition separating the two fixed points at some finite ratio of the quadratic to quartic couplings. However, in this case the quadratic couplings are  relevant and always lead to a free fixed point in the low-energy limit. Hence, in contrast to the model we propose, such a system would not exhibit a quantum critical point.

Before proceeding we mention that other genralizations of the SYK model were introduced to explore different aspects of the physics \cite{Gu2016,Berkooz2016,Gross2016}. For example Gu and Qi \cite{Gu2016} considered a chain of SYK sites coupled through local quartic interactions, which allows to study the relation between transport coefficients and the Butterfly effect in extended systems. Our goal with the model we introduce  is different. It is to allow tuning of a  quantum critical point separating phases with distinct chaotic behavior.

The rest of the paper is organized as follows. In Sec.\ref{sec.GreenFunction} we discuss the coupled self consistent equations for the Green's functions of the SYK fermions and the peripheral fermions. 
From the solution we identify the critical point separating the SYK like phase from the weakly coupled Fermi liquid. In Sec.\ref{sec.LowTEntropy} we compute the $T=0$ entropy of the model, showing how it vanishes continuously at the critical point. The weak coupling phase, unlike the SYK-like phase has vanishing zero-point entropy. In Sec.\ref{sec.OTOC} we turn to compute four-point out-of-time-order (OTO) correlation functions, which encapsulate the scrambling dynamics. The results are discussed and summarized in Sec.\ref{sec.Conclusion}. Some details of the calculations and numerical computations are given in the Appendices \ref{app.GcSYK},  \ref{app.LuttingerTheorem}, \ref{app.OTOC} and in the Supplementary Information.

\section{Two-point function and the quantum critical point} \label{sec.GreenFunction}
\begin{center}
\begin{figure}
\includegraphics[width=0.5\textwidth]{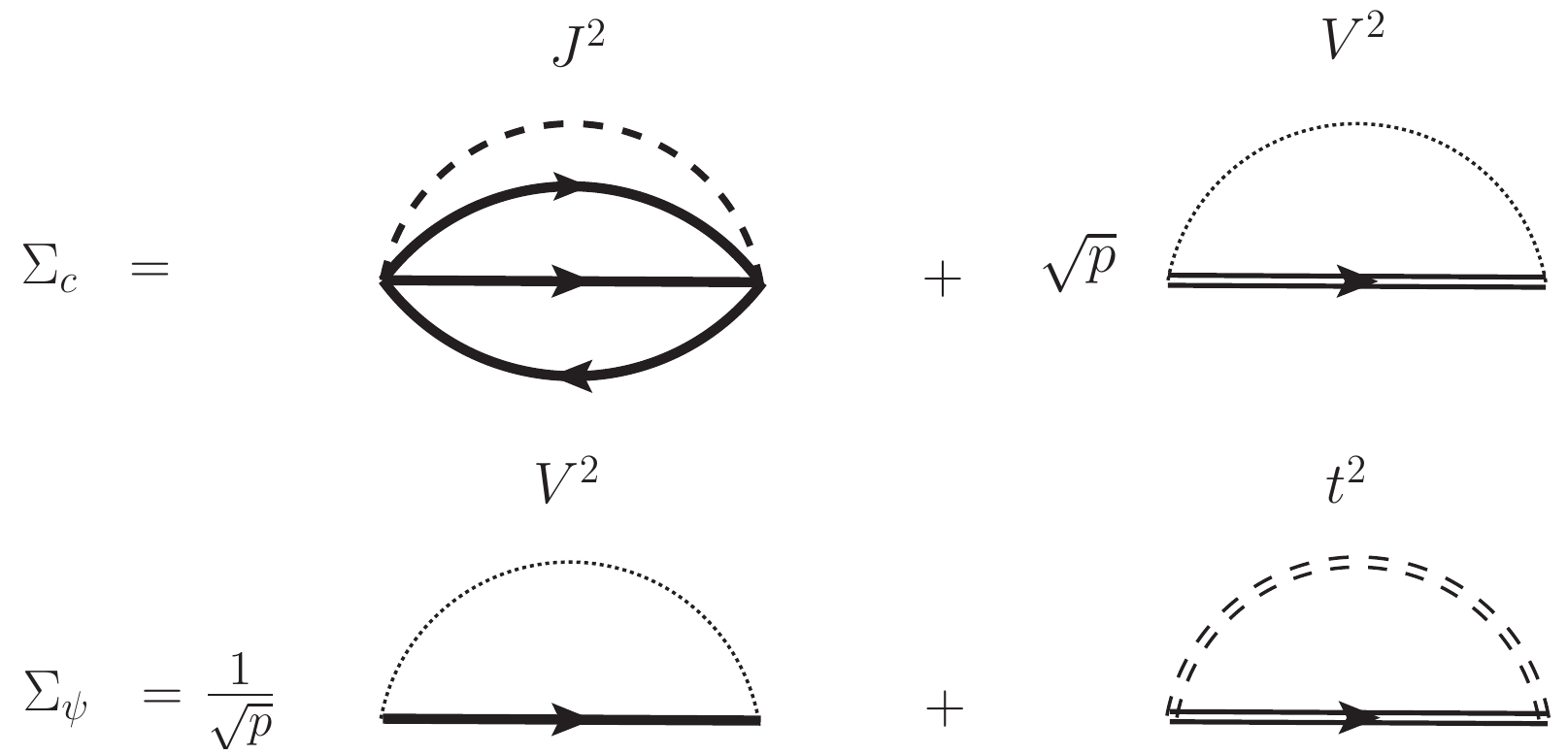}
\caption{Self-energy diagrams that contribute at leading order
in $1/N$ for a fixed ratio $p=M/N$. The bold lines represent Green's function $G$ of the SYK sites and the double line the Green's function $\mc{G}$ for the peripheral sites. The dashed, dotted and dashed-double lines imply disorder averaging over $J_{ijkl}$, $V_{i\a}$ and $t_{\a\b}$, respectively.}
\label{fig.SelfEnergy}
\end{figure}
\end{center}

In this section we discuss the Green's functions for the two species of the fermions and show how the quantum phase transition as a function $p$ is manifested in the single-particle spectral properties.
Here we are interested in the disorder averaged Green's functions, $G(\tau)=-\overline{\la\mc{T}_\tau c(\tau)c^\dag(0)\ra}$ and $\mc{G}(\tau)=-\overline{\la\mc{T}_\tau \psi(\tau)\psi^\dag(0)\ra}$, where $\tau$ is the imaginary time and the over-line in $\overline{\la\dots\ra}$ denotes averaging over realizations of $\{J_{ijkl},t_{\a\b},V_{i\a}\}$. For $N\to\infty$  the Green's functions can be obtained either diagrammatically or, equivalently, from the saddle point of an effective action functional obtained via the replica formalism (see Supplementary Information). In either way we obtain the following self-consistent Schwinger-Dyson equations
\begin{subequations}\label{eq:SelfConsistency}
\begin{align}
G^{-1}(i\omega_{n})= & i\omega_{n}+\mu-\Sigma_{J}(i\omega_{n})-V^{2}\sqrt{p}\mathcal{G}(i\omega_{n})\label{eq:SelfConsistency1}\\
\mathcal{G}^{-1}(i\omega_{n})= & i\omega_{n}+\mu-\frac{V^{2}}{\sqrt{p}}G(i\omega_{n})-t^{2}\mathcal{G}(i\omega_{n})\label{eq:SelfConsistency2}
\end{align}
where $\w_n=(2n+1)\pi T$ is the fermionic Matsubara frequency and $n$ an integer. The last two terms of each of the equations above correspond to the self-energy diagrams of Fig.\ref{fig.SelfEnergy}. Due to the large $N$ limit and the disorder averaging, only the `rainbow' diagram (Fig.\ref{fig.SelfEnergy}), with the bare Green's function lines replaced by the dressed ones, contributes to the interaction correction to the self-energy at leading order in $1/N$ \cite{KitaevKITP,Polchinski2016}, i.e. 
\begin{align}
\Sigma_{J}(\tau)= & -J^{2}G^{2}(\tau)G(-\tau)\label{eq:SelfConsistency3}
\end{align} 
\end{subequations}
As in Refs.\onlinecite{Sachdev1993,Sachdev2015} we define the shifted self energy $\hat{\Sigma}(i\omega)=\Sigma(i\omega)-\mu$ in order to eliminate the chemical potential from the equations. This shift, only affects the  behavior at $\tau=0$ and therefore does not affect Eq. (\ref{eq:SelfConsistency}c) in the long time limit we are interested in.

Below we show that the coupled self-consistency equations equations \eqref{eq:SelfConsistency} lead to two distinct phases at $T=0$, tuned by the ratio $p$. Both phases enjoy an emergent low-energy conformal symmetry, but with distinct scaling dimensions of the two-point functions. %To this end, we consider the self-consistency equations for the retarded Green's functions $G_R(\w)$, $\mc{G}_R(\w)$, obtained by analytically continuing equations\eqref{eq:SelfConsistency} to real frequency $\w$ via $i\w_n\to\w+i\eta$.

\subsection{Non-Fermi liquid} \label{subsec.NFL}
\begin{center}
\begin{figure*}
\begin{tabular}{cc}
\includegraphics[width=0.5\textwidth]{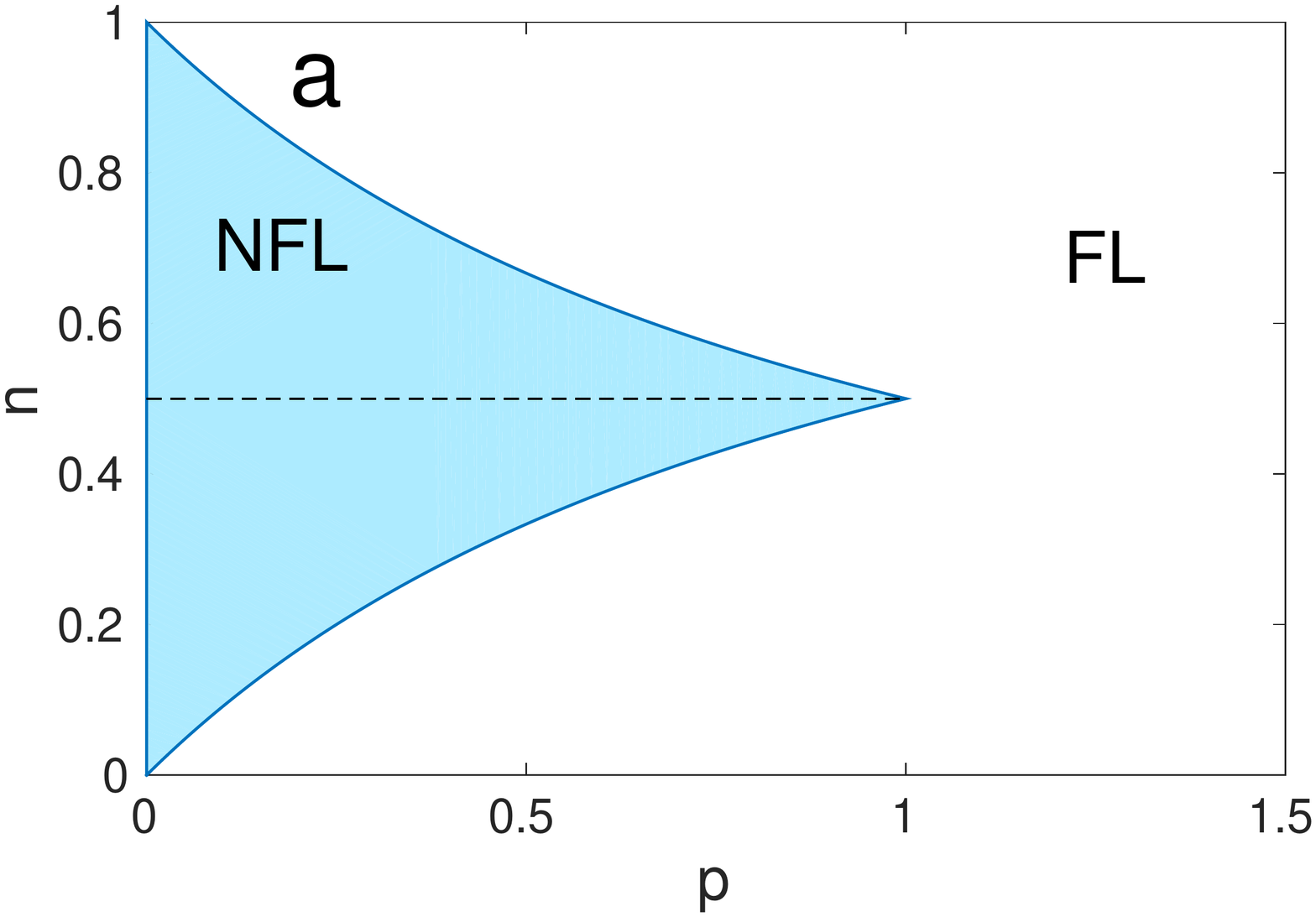}&
\includegraphics[width=0.5\textwidth]{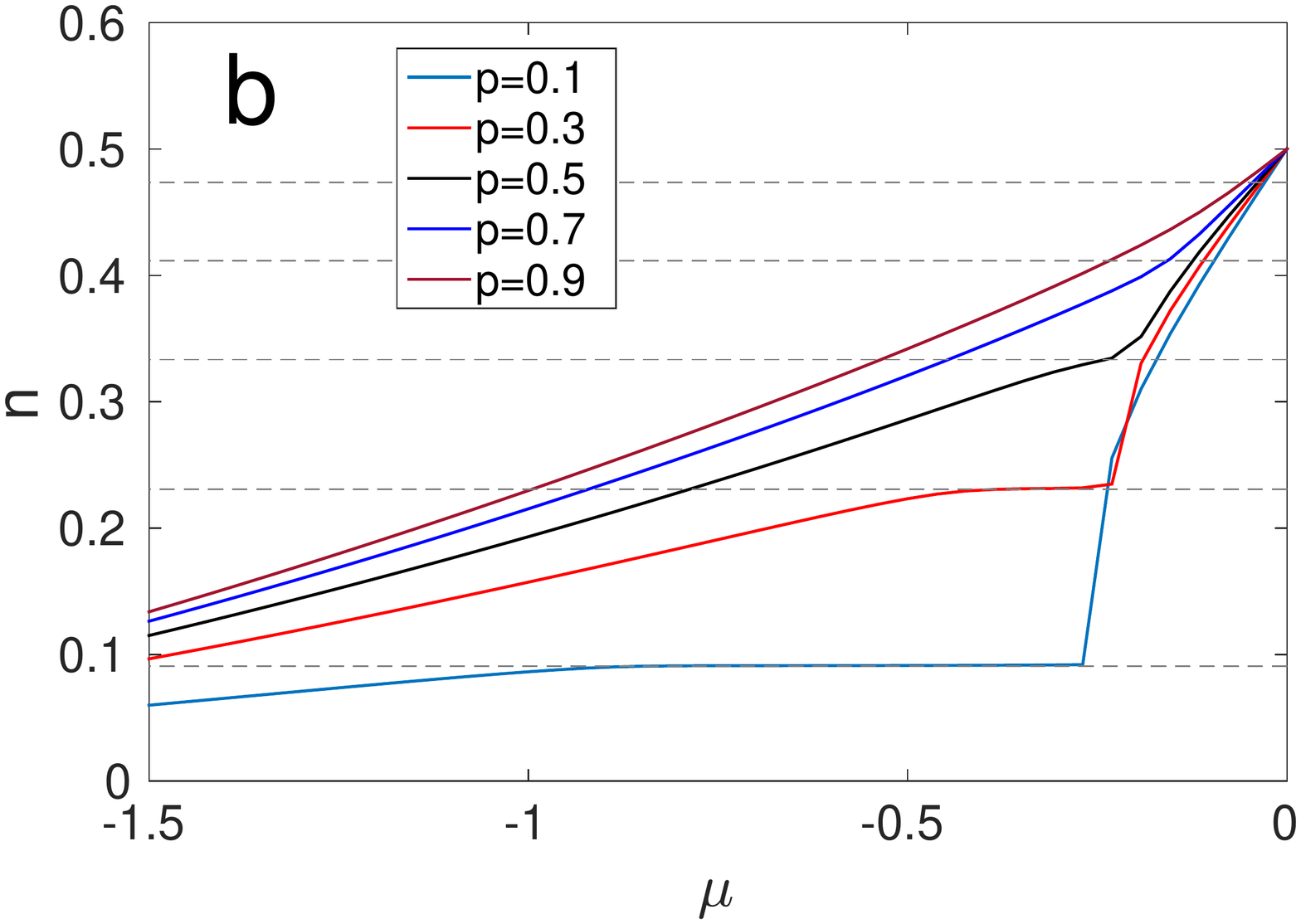}
\end{tabular}
\caption{(a) Phase diagram of the generalized SYK model in the plane of the average fermion filling $n$ and the ratio $p=M/N$. The dashed line indicates half filling. (b) Fermion density $n$, below half filling, as a function of the chemical potential $\mu$ for $p=0.1,\dots,0.9$ (bottom to top curve) at $T=0.025J$. The dashed lines indicate respective critical densities $n_c$'s at the lower phase boundary in panel (a). The plateaus for $p=0.1,0.3$ imply the presence of incompressible state at $n_c$.}
\label{fig.PhaseDiagram}
\end{figure*}
\end{center}

The solution for $p<p_c$ is found by first neglecting the terms $i\w_n$ and $t^2\mc{G}(i\w_n)$ in  equations \eqref{eq:SelfConsistency}, with the expectation that the solution will justify this omission in the low-frequency limit.
Without the omitted terms the equations assume the following simple form in imaginary time:
\begin{subequations} \label{eq.SYKSaddle}
\bea
&&\int_0^\b d\tau_1 G(\tau,\tau_1)[\hat{\S}_J(\tau_1,\tau')+V^2\sqrt{p}\mc{G}(\tau_1,\tau')]=-\delta(\tau-\tau')\nn\\
&&\\
&&\frac{V^2}{\sqrt{p}}\int_0^\b d\tau_1 \mc{G}(\tau,\tau_1)G(\tau_1,\tau')=-\delta(\tau-\tau') 
\eea
\end{subequations}
As in the pure SYK model, these equations along with  equation \eqref{eq:SelfConsistency3} are invariant under arbitrary reparameterization of imaginary time $\tau=f(\s)$ with the following scaling of Green's functions and self-energy $\hat{\S}_J$,
\begin{subequations} \label{eq.SYKConformalSymmetry}
\bea
\tilde{G}(\sigma,\sigma')&=&[f'(\sigma)f'(\sigma')]^{\D_c} G(f(\s),f(\s'))\frac{g(\s')}{g(\s)} \\
\tilde{\mc{G}}(\sigma,\sigma')&=&[f'(\sigma)f'(\sigma')]^{\D_\psi}\mc{G}(f(\s),f(\s'))\frac{g(\s')}{g(\s)} \\
\tilde{\Sigma}_J(\sigma,\sigma')&=&[f'(\sigma)f'(\sigma')]^{\D_\S}\hat{\S}_J(f(\s),f(\s'))\frac{g(\s')}{g(\s)}
\eea
\end{subequations}
with $f'(\s)=\partial f/\partial \s$.  Here the  scaling dimension for SYK fermions is $\D_c=1/4$, the peripheral fermions have the scaling dimension $\D_\psi=3/4$, and $\D_\S=3/4$. The factor $g(\s)$, real in imaginary time, is due to an additional emergent $U(1)$ gauge symmetry for the complex fermions as discussed in Ref.\onlinecite{Sachdev2015}. 

The conformal symmetry of the equations leads to solutions with power-law forms 
\begin{subequations}\label{eq.SYKsolution}
\bea 
G_R(\w)&=&\Lambda\frac{e^{-i(\pi/4+\theta)}}{\sqrt{J\w}} \label{eq.SYKG}\\
\mc{G}_R(\w)&=&-\frac{\sqrt{p}}{V^2\Lambda}e^{i(\pi/4+\theta)}\sqrt{J\w} \label{eq.SYKg} \\
\hat{\S}_J^R(\w)&=&-\pi^{-1}\L^3 e^{i(\pi/4+\t)}\cos2\t \sqrt{J\w} \label{eq.SYKSigma} 
\eea
\end{subequations}
where we have performed the analytic continuation $i\w_n\to\w+i\eta$ to obtain the retarded Green's functions. The constant $\Lambda$, determined by direct substitution of the power-law forms into the conformal self consistency equations  \eqref{eq.SYKSaddle}, is given by 
\bea\label{Lambda}
\L&=&\left(\frac{(1-p)\pi}{\cos2\theta}\right)^{1/4}
\eea 

The parameter $\t$ in equations \eqref{eq.SYKsolution} is related to spectral asymmetry and fermion filling  through a Luttinger theorem\cite{Sachdev1993,Sachdev2015}. We show in  Appendix \ref{app.LuttingerTheorem} that for our model in the NFL fixed point the Luttinger relation takes the form
\begin{align}
n= & \frac{1}{1+p}\left[\left(\frac{1}{2}-\frac{\theta}{\pi}\right)+p\left(\frac{1}{2}+\frac{\theta}{\pi}\right)-(1-p)\frac{\sin2\theta}{4}\right],\label{eq:LuttingerTheorem}
\end{align}
The derivation of this Luttinger relation only uses information about $G(\omega)$, $\mathcal{G}(\omega)$ and $\Sigma_{J}(\omega)$ for $\omega\to0$, known from the conformal limit, and $\omega\to\infty$, determined by fermion anticommutation. Hence, the relation does not depend on cutoff or any other parameter of the model like $V$, $t$ or $J$.  To ensure $-\mrm{Im}G_R(\w),-\mrm{Im}\mc{G}_R(\w)>0$, $\theta$ is restricted to the range $[-\pi/4,\pi/4]$, which through the above Luttinger relation determines the range of densities over which the  solutions \eqref{eq.SYKsolution} exist, namely
\begin{align}
\frac{p}{1+p}\leq & n\leq\frac{1}{1+p}\label{eq:DensityRange}
\end{align}
This defines a region on the $p-n$ plane, shown in Fig.\ref{fig.PhaseDiagram}(a), in which the NFL fixed-point is stable. The phase $\theta$ changes continuously as the density is varied between the lower phase boundary of the NFL, where $\theta=\pi/4$ to the upper boundary $\theta=-\pi/4$. $\theta=0$ corresponds to the particle hole symmetric line at half filling, where the model of equation \eqref{eq.Model} is essentially equivalent to a model of Majorana fermions discussed in section \ref{sec.OTOC.majorana}.  

We should further verify that the conformal Green's functions \eqref{eq.SYKsolution}  solve the full self-consistency equations \eqref{eq:SelfConsistency} at low frequency by directly substituting them in the full equations. This shows that the terms omitted from equations \eqref{eq:SelfConsistency} to obtain equation \eqref{eq.SYKSaddle}  become negligible below a cutoff scale $\w_c$ (see Appendix \ref{app.GcSYK}). For example at half filling $\w_c\propto (V^4/t^2J)(\sqrt{1-p}/p)$ for $p\to 1$. Hence this characteristic frequency scale associated with emergence of conformal symmetry vanishes continuously at a critical point $p_c=1$ (at half filling). It is interesting to note that the location of the critical point depends only on $n$ and $p$ regardless of the coupling strength $V$ between the two species. However, as mentioned above, the coupling $V$ does control the frequency scale at which the low energy fixed point emerges. 

A defining feature of the NFL phase of this model, both at half filling and away from half filling, is the singularity at $\w\to 0$ in the single particle spectral functions. The density of states (DOS) of the $c$ fermions behaves as $1/\sqrt{\w}$, as found in the original model by Sachdev and Ye \cite{Sachdev1993}, whereas that of the $\psi$ fermions is suppressed at low frequency as $\sqrt{\omega}$. 
The latter is similar to the well-known zero-bias suppression due to the combined effect of interaction and disorder \cite{Altshuler1985}.
The constant $\L$,  which determines the strength of the low frequency singularity, has a singular behavior as the system is tuned toward the phase transition $p\to p_c$ (or $n\to n_c$ if the density is used as a tuning parameter). 

On the particle hole symmetric line the strength of the $1/\sqrt{\w}$ peak in $G_R$ vanishes as $(1-p)^{1/4}$. At the same time the strength of the $\sqrt{\w}$ singularity in the DOS of the peripheral fermions diverges as  $(1-p)^{-1/4}$. The evolution of the low frequency singularity is quite different when approaching the transition near top and bottom phase boundaries, away from half filling. There, as $\theta\to \pm\pi/4$, the singularities acquire a strong asymmetry  between positive or negative frequencies. For example, near the lower phase boundary we take $\theta=\pi/4-\delta\theta$ to find the leading behavior of the spectral functions in the small detuning parameter $\delta\theta$. The strength of the $1/\sqrt{\w}$ singularity in the DOS of the $c$ fermions diverges as $1/\delta\theta^{1/4}$ for $\w>0$, but vanishes as $\delta\theta^{3/4}$ for $\w<0$, whereas the DOS of the peripheral $\psi$ fermions vanishes as $\d\theta^{1/4}$ for $\w>0$ and as $\d\theta^{5/4}$ for $\w<0$. At the same time, the cuoff for the conformal behaviour collapses approaching the critical point as $\sim J\d\theta^{1/2}$ for positive frequencies and as $\sim J\d\theta^{5/2}$ for negative frequencies.

The vanishing of the spectral functions for $\w\to 0^\pm$ on the upper (lower) boundaries may indicate a phase transition into an incompressible state. To assess this possibility we solved the self consistency equations \eqref{eq:SelfConsistency} numerically and obtained the fermion density as a function of the chemical potential at a fixed value of $p$.  The results displayed in Fig. \ref{fig.PhaseDiagram}(b) indicate a transition to an incompressible state, seen as a plateau in the density. This incompressible state appears as a line in the canonical phase diagram (part of the top and bottom phase boundaries), but it covers a non-vanishing area in the grand-canonical phase diagram  $\mu$ versus $p$. For values of $p$ closer to 1, we find a direct transition to a metallic (compressible) Fermi liquid as is also seen in the Fig. \ref{fig.PhaseDiagram}(b). See the Supplementary Information for more details of the transition away from half filling.

Our discussion so far pertained only to the $T=0$ Green's functions. Later, for the sake of calculating the out of time order correlation functions we will need the analytic continuation to low non-vanishing temperature. These Green's functions are obtained from the $T=0$ solutions \eqref{eq.SYKsolution} using the conformal symmetry as shown in Appendix \ref{app.GcSYK}.

\begin{center}
\begin{figure*}
\includegraphics[width=\textwidth]{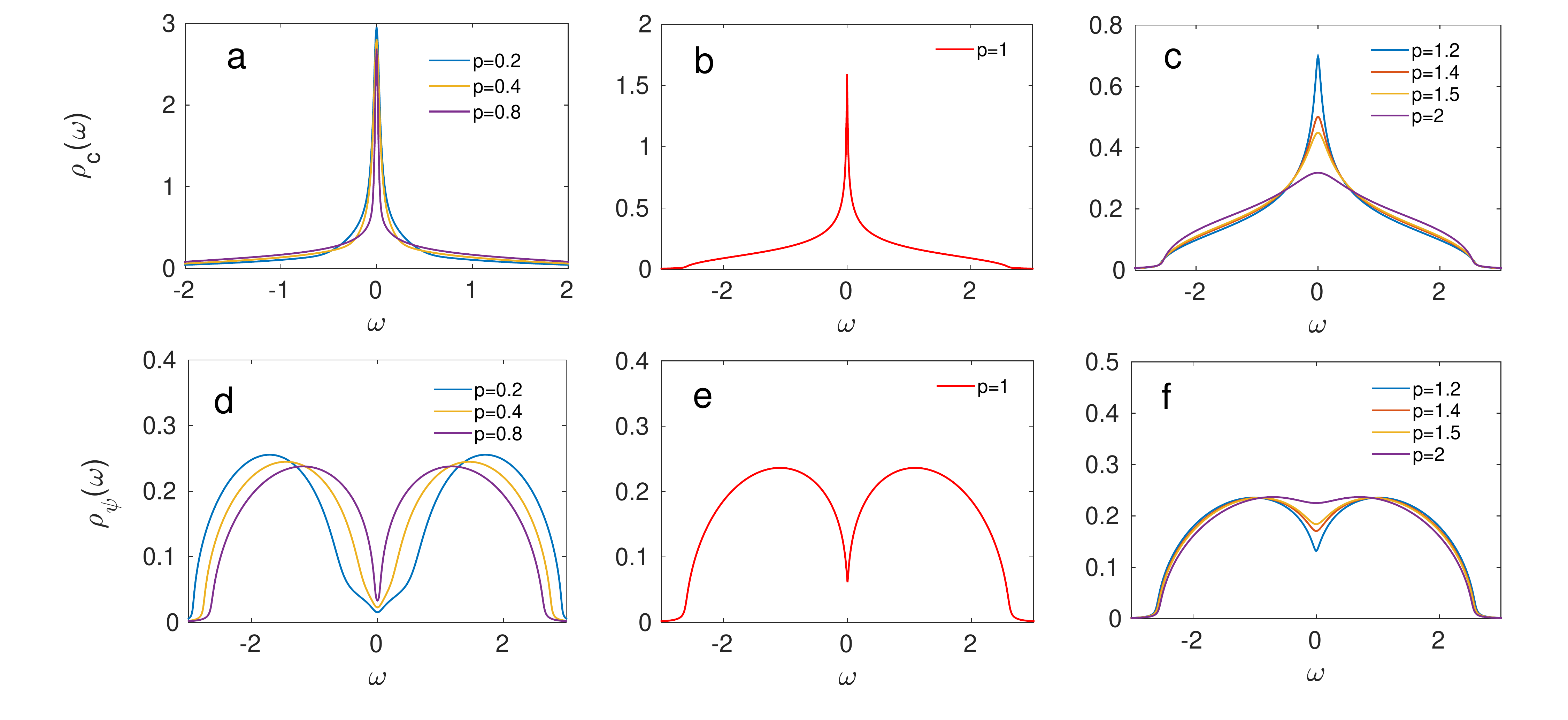}
\caption{Numerical results for the single particle spectral function on the full frequency range taken on on the SYK sites [panels (a) to (c)] and the quadratic sites [panels (d) to (f)]. The spectral functions are computed at $T=0.025J$ and varying values of $p$ showing how the low temperature singularities vanish beyond the critical point.}
\label{fig.SpectralFunction}
\end{figure*}
\end{center}

\subsection{Fermi liquid} \label{subsec.FL}
A different $T=0$ solution of the self-consistency equations  [equations\eqref{eq:SelfConsistency}] emerges for $p>p_c(n)$. 
In this regime we first solve the equations \eqref{eq:SelfConsistency} by neglecting the interaction self energy $\Sigma_J(i \w_n)$ as well as the free $i\w_n$ terms. In the particle hole symmetric case $\mu=0$  ($n=1/2$), we obtain the following Green's functions
\begin{subequations} \label{eq.FLSolution}
\bea
G_R(\w)&=&-i\frac{1}{\sqrt{p(p-1)}}\frac{t}{V^2}\\
\mc{G}_R(\w)&=&-i\sqrt{\frac{p-1}{p}}\frac{1}{t}
\eea 
\end{subequations}
The above is a valid physical solution of the full equations for $p>p_c=1$ and frequencies much lower than an emergent cutoff scale %$\w_0\propto t\sqrt{p/(p-1)}$ for $p\gg1$ and 
$\w_0\propto (V^2/t)\sqrt{p(p-1)}$ that vanishes at the critical point $p=1$. Also on approaching $p_c=1$ from above, the low-frequency DOS of the SYK sites diverges as $(p-1)^{-1/2}$ and the DOS of the peripheral sites vanishes as $(p-1)^{1/2}$, continuously merging with the singularities $G(\w)\sim 1/\sqrt{\w}$ and  $\mc{G}(\w)\sim \sqrt{\w}$ respectively on the other side of the transition  ($p<p_c$). 

To obtain the low energy Fermi liquid solution \eqref{eq.FLSolution} we have omitted the self energy $\S_J(i\w)$ from the self consistency equations  \eqref{eq:SelfConsistency}. We can now feed the solutions back to a calculation of the low-frequency behavior of $\Sigma_J$. The result will be valid for $\w\ll \w_0$. Our starting point for this calculation is the Fourier transformed Eq.\eqref{eq:SelfConsistency3}
 \begin{align}
&\S_J(i\w_n)=\int_0^\b d\tau e^{i\w_n\tau}\S(\tau)\nn\\
&=-\frac{J^2}{\b^2}\sum_{n_1,n_2}G(i\w_{n_1})G(i\w_{n_2})G(i\w_{n_1}+i\w_{n_2}-i\w_n) 
\end{align}
Carrying out the Matsubara summations and the analytical continuation $i\w_n\to\w+i\eta$ gives (see Supplementary Information)
%\begin{align}
%&\mrm{Im}\S_J^R(\w)=J^2\int \prod_{i=1}^3\left(d\w_i \rho_c(\w_i)\right)\left[n_\mrm{F}(-\w_1)n_\mrm{F}(\w_2)n_\mrm{F}(-\w_3)\right.\nn\\
%&\left.+n_\mrm{F}(\w_1)n_\mrm{F}(-\w_2)n_\mrm{F}(\w_3)\right]\d(\w-\w_1+\w_2-\w_3)
%\end{align}
%or, at $T=0$ 
\begin{align}
&\mrm{Im}\S_J^R(\w>0)\nn\\
&=-J^2\pi\int_0^{\w_1+\w_2\leq \w} d\w_1d\w_2\rho_c(\w_1)\rho_c(\w_2)\rho_c(\w-\w_1-\w_2)
\end{align}
for $T=0$. In the low frequency limit, $\w\ll\w_0$, the integral can be evaluated using the constant DOS $\rho_c=-(1/\pi)\mrm{Im}G_R(\w)=W/(\pi\sqrt{p(p-1)}g^2)$, leading to 
\bea\label{eq.SigmaFL}
\mrm{Im}\S_J^R(\w)&\approx & -\frac{J^2t^3}{2\pi^2V^6}\frac{1}{(p(p-1))^{3/2}}\w^2 
\eea
Hence, the $T=0$ quasiparticle decay rate vanishes as $\w^2$ for $\w\to 0$, as expected for a Fermi liquid. At the same time the pre-factor of the $\w^2$ dependence diverges as $(p-1)^{-3/2}$ for $p\to p_c=1$ indicating the breakdown of the Fermi-liquid at the critical point.
In the Fermi liquid phase, the saddle-point equations [equations\eqref{eq:SelfConsistency}] have a trivial emergent conformal symmetry with scaling dimensions $\D_c=\D_\psi=1/2$, corresponding to non-interacting fermions.

 Hence the model of equation \eqref{eq.Model} is an example of a solvable model for NFL-FL transition.  The transition has some similarity with the overscreeing to underscreening transition in multichannel $SU(N)$ Kondo impurity model \cite{Parcollet1997}. However, in our case the NFL-FL transition is realized in a much simpler setting.

\subsection{Numerical results for spectral function across the QCP}\label{subsec.NumGreenFunction}
To corroborate the analytical results discussed in the preceding subsections, we have solved the self-consistency equations [equations\eqref{eq:SelfConsistency}] numerically for the retarded Green's functions $G_R(\w)$ and $\mc{G}_R(\w)$ over a range of $p$ across the transition. The numerical calculation is performed at finite temperatures. 

Here we discuss the results for half filling, $\mu=0$. The results away from half filling are discussed in the Supplementary Information.
In Fig.\ref{fig.SpectralFunction}, we show the evolution of spectral functions, $\rho_c(\w)=-(1/\pi)\mrm{Im}G_R(\w)$ and $\rho_\psi(\w)=-(1/\pi)\mrm{Im}\mc{G}_R(\w)$, with $p$ for the two species of fermions at $T=0.025J$ and $t=V=J$. In the NFL phase [Figs.\ref{fig.SpectralFunction}(a),(d)], the spectral functions match at low energies with the finite-temperature spectral densities obtained in the conformal limit, showing in particular the $1/\sqrt{\w}$ and $\sqrt{\w}$ singularities in $G_R(\w)$ and $\mc{G}_R(\w)$ respectively. Upon crossing to the FL side of the transition  these singularities are rounded off even at $T=0$. This can be clearly seen deep inside the FL phase for $p=1.5,2$ [Figs.\ref{fig.SpectralFunction}(c),(f)]. 

In the next section, we show that the NFL-FL transition also manifests itself in an intriguing way in the evolution of low-temperature entropy.

\section{Zero-temperature entropy} \label{sec.LowTEntropy}
\begin{center}
\begin{figure}
\includegraphics[width=0.5\textwidth]{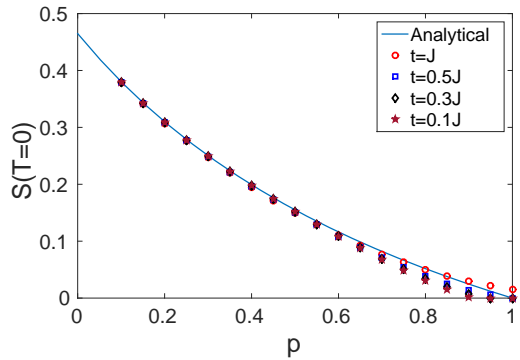}
\caption{The zero temperature limit of the entropy as a function of $p$. The analytic result using the Luttinger theorem and an assumption $S(T\to 0)=0$ at the finite density phase transition is compared to a direct numerical evaluation of the
low-temperature entropy and extrapolations to $T = 0$ for various values of $t$.}
\label{fig.ZeroTEntropy}
\end{figure}
\end{center}

The SYK model has an extensive zero-temperature residual entropy \cite{KitaevKITP,Sachdev2015,Fu2016}, when the thermodynamic limit $N\to \infty$ is taken before taking the zero temperature limit. The residual entropy stems from a dense many-body energy  spectrum with level spacing $\sim e^{-N}$ \cite{Fu2016,Maldacena2015}, even near the ground state. In this sense, the low-energy spectrum of the SYK Hamiltonian resembles the spectrum of usual quantum many-body systems at high energies.   

To derive the residual entropy we adopt the scheme outlined in Ref.\onlinecite{Sachdev2015}. The idea is to find the derivative $(\partial S/\partial n)_{T=0}$ as a function of the density $n$, then perform the integration to the desired density in order to obtain $S_0$. The derivative is found using the Maxwell relation
\begin{align}
\left(\frac{\partial S}{\partial n}\right)_{T}= & -\left(\frac{\partial\mu}{\partial T}\right)_{n},
\end{align} 
The right hand side has been related in Refs.\onlinecite{Parcollet1998,Sachdev2015} to the spectral asymmetry affected by $\theta$. Specifically  $(\partial\mu/\partial T)_n=-\ln(\tan(\pi/4+\theta))+O(T^{2\Delta_c})$. The derivation of this relation is exactly the same here as in the SYK model \cite{Sachdev2015}.
 
Now we can use the Luttinger relation \eqref{eq:LuttingerTheorem}, connecting $\theta$ to the density, to compute the entropy as an integral over $\theta$. 
\begin{align}
S_0(n)= & S_0(n_0)+\int_{n_0}^{n}dn \ln(\tan(\pi/4+\theta(n))) \label{eq.nIntegration}
\end{align}
Here $n_0$ refers to a reference density at which the $T=0$ entropy is
known. For example, at $p=0$, for the pure SYK model, the natural
choices are either the empty state, $n_0=0$, or the completely filled state, $n_0=1$, corresponding to $\theta=\pm\pi/4$, respectively. In these cases we expect $S_0=0$ \cite{Sachdev2015}. For general $p$, this is not as straightforward since the NFL phase only exists up to a lower (upper) critical density $n_c=p/(p+1)>0$ ($n_c=(p+1)^{-1}<1$). 
However, assuming vanishing of the residual entropy at  $n_c$ (upper and lower phase boundaries), that is  $S_0(n_0(\t=\pm\pi/4))=0$, we can take one of the boundaries, e.g. $\t=\pi/4$, as the reference state to obtain
\begin{align}
&S_0(n(\theta))= \frac{1-p}{1+p}\int_{\pi/4}^{\theta}d\theta\ln(\cot(\pi/4+\theta))\left(\frac{1}{\pi}+\frac{\cos2\theta}{2}\right)
\end{align} 
At half filling ($\t=0$), we get
\begin{align}\label{eq.ZeroTEntropy}
S_0(n=1/2)= & \frac{1-p}{1+p}S_{SYK}(T=0)
\end{align}
 Where $S_{SYK}(T=0)\simeq0.46$ \cite{Fu2016,Georges2000}, is the residual entropy of the SYK
model. Hence the $T=0$ entropy at half filling vanishes continuously at the QCP as $(1-p)$. 
At fixed $p<1$ the entropy vanishes upon approaching the upper and lower phase as $\delta S\sim |n-n_c| \log(1/|n-n_c|)$. 

We corroborate the analytic result for the residual entropy at half filling with a numerical calculation of $S(T)$, extrapolated to $T=0$ using the linear in $T$ behavior of the entropy in the low temperature conformal limit (see the Supplementary Information). Note that this extrapolation is hindered close to the critical point due to collapse of the cutoff $\omega_c$ beyond which the entropy is no longer linear in $T$, hence the numerical result is not accurate in this regime. The numerically calculated values of the zero temperature entropy are shown in Fig. \ref{fig.ZeroTEntropy} and compared to the analytic curve. 

Vanishing of the entropy at the critical point suggests a fundamental change of the geometry in the dual gravity picture of the transition, which involves elimination of the black hole. Below we give further evidence to this view from the perspective of the scrambling dynamics . We confirm that the NFL to FL critical point marks a transition in the nature of many-body quantum chaos.

\section{Out-of-time-ordered correlations: scrambling} \label{sec.OTOC}

Following Maldacena, Shenker and Stanford \cite{Maldacena2015}, we characterize the scrambling dynamics through a Lyapunov exponent $\lambda_\mrm{L}$ characterizing the four-point out-of-time-ordered (OTO) correlation functions.  We find, $\lambda_\mrm{L}\to 2\pi T$, saturating the chaos bound \cite{Maldacena2015} in the conformal low temperature limit, over the entire NFL phase in Fig.\ref{fig.PhaseDiagram}(a). 

\subsection{Majorana fermion model}\label{sec.OTOC.majorana}
Before computing  the OTO correlations in the model \eqref{eq.Model}, we treat a closely related model of Majorana fermions, which admits a much simpler calculation. The generalization to complex fermions at arbitrary filling will follow in subsection \ref{sec.OTOC.complex}.
%To simplify the presentation of the diagramatic calculation involved, we study a Majorana fermion model equivalent to the model of equation \eqref{eq.Model} at half filling. 
The model consists of two species of Majorana fermions: $\chi_i$ in place of the complex fermion $c_i$ on sites $i=1,\dots,N$ and $\eta_\a$ in place of the fermion $\psi_\a$ on sites $\a=1,\dots,M$ (see Fig.\ref{fig.Model}). Specifically, we consider the following Hamiltonian:
 \begin{align}\label{eq.MajoranaModel}
\mathcal{H}= & \frac{1}{4!}\sum_{ijkl}J_{ijkl}\chi_{i}\chi_{j}\chi_{k}\chi_{l}+\frac{i}{2!}\sum_{\alpha\beta}t_{\alpha\beta}\eta_{\alpha}\eta_{\beta}+i\sum V_{i\alpha}\chi_{i}\eta_{\alpha}
\end{align}
 where $J_{ijkl},~t_{\a\b},~V_{i\a}$ are all real; $J_{ijkl}$ and $t_{\alpha\beta}$ are fully antisymmetric
and $\langle J_{ijkl}^{2}\rangle=J^{2}3!/N^{3}$, $\langle t_{\alpha\beta}^{2}\rangle=t^{2}/M$
and $\langle V_{i\alpha}^{2}\rangle=V^{2}/\sqrt{NM}$ with $p=M/N$ as in the complex fermion case. The model leads to same large-$N$ saddle-point equations as in \eqref{eq:SelfConsistency} with $\mu=0$, for the Green's functions $G(\tau)=-\overline{\la\mc{T}_\tau\chi_i(\tau)\chi_i(0)\ra}$ and $\mc{G}(\tau)=-\overline{\la\mc{T}_\tau\eta_\a(\tau)\eta_\a(0)\ra}$. At $p=0$, this reduces to the version of the SYK model proposed by Kitaev \cite{KitaevKITP}. 

The out-of-time-order (OTO) correlations used to diagnose quantum chaos involve four Majorana operators. For example the OTO correlation on two SYK sites is expected to take the form,     
\begin{align}
\overline{\la\chi_i(t)\chi_j(0)\chi_i(t)\chi_j(0)\ra}\simeq f_0-\frac{f_1}{N}e^{\l_\mrm{L}t}+\mc{O}\left(\frac{1}{N^2}\right), \label{eq.OTOCKitaev}
\end{align}
and expected to hold up to some intermediate time scale $t \aplt t^*\simeq (1/\l_\mrm{L})\ln(N)$, called the scrambling time. This is the time over which the OTO correlation decays to small values and information encoded in local observables is lost to operators encompassing the  entire system\cite{Maldacena2015}. Here $\l_\mrm{L}$ is the Lyapunov exponent \cite{Larkin1969}, or scrambling rate, which obeys a universal upper bound $\l_\mrm{L}\le 2\pi T$ \citep{Maldacena2015}. %The latter paper also introduced the modified OTO correlations in which, for computational convenience the four, operators are rotated, from each other by $1/4$ of the thermal circle. 

In our model \eqref{eq.MajoranaModel}, due to the coupling between the SYK fermions $\chi_i$ and the peripheral sites $\eta_\a$  the OTO correlation \eqref{eq.OTOCKitaev} cannot be found independently of the correlation function describing ``cross scrambling" of the SYK sites with the peripheral fermions. Specifically, we compute the following two coupled four-point functions, $F_{\chi\chi\chi\chi}=F_1$ and $F_{\eta\eta\chi\chi}=F_2$,
\begin{subequations} \label{eq.OTOC}
\begin{align}
F_1(t_{1},t_{2})&= \frac{1}{N^{2}}\sum_{ij}\overline{\mrm{Tr}[y\chi_{i}(t_{1})y\chi_{j}(0)y\chi_{i}(t_{2})y\chi_{j}(0)]} \label{eq:OTOC1}\\
F_2(t_{1},t_{2})= & \frac{1}{NM}\sum_{i\alpha}\overline{\mrm{Tr}[y\eta_{\alpha}(t_{1})y\chi_{i}(0)y\eta_{\alpha}(t_{2})y\chi_{i}(0)]}.
\label{eq:OTOC2}
\end{align}
\end{subequations}
Here we  used a modified version of the OTO correlations, in which the four operators are rotated from each other by $1/4$ of the thermal circle, i.e.  $y^4=e^{-\b\mc{H}}/Z$. This modified OTO correlation was introduced in Ref.\onlinecite{Maldacena2015} for  computational convenience. The operator $y$ helps to regularize the four point function in the conformal limit \cite{Maldacena2015,Maldacena2016}. 

\begin{center}
\begin{figure}
\includegraphics[width=0.5\textwidth]{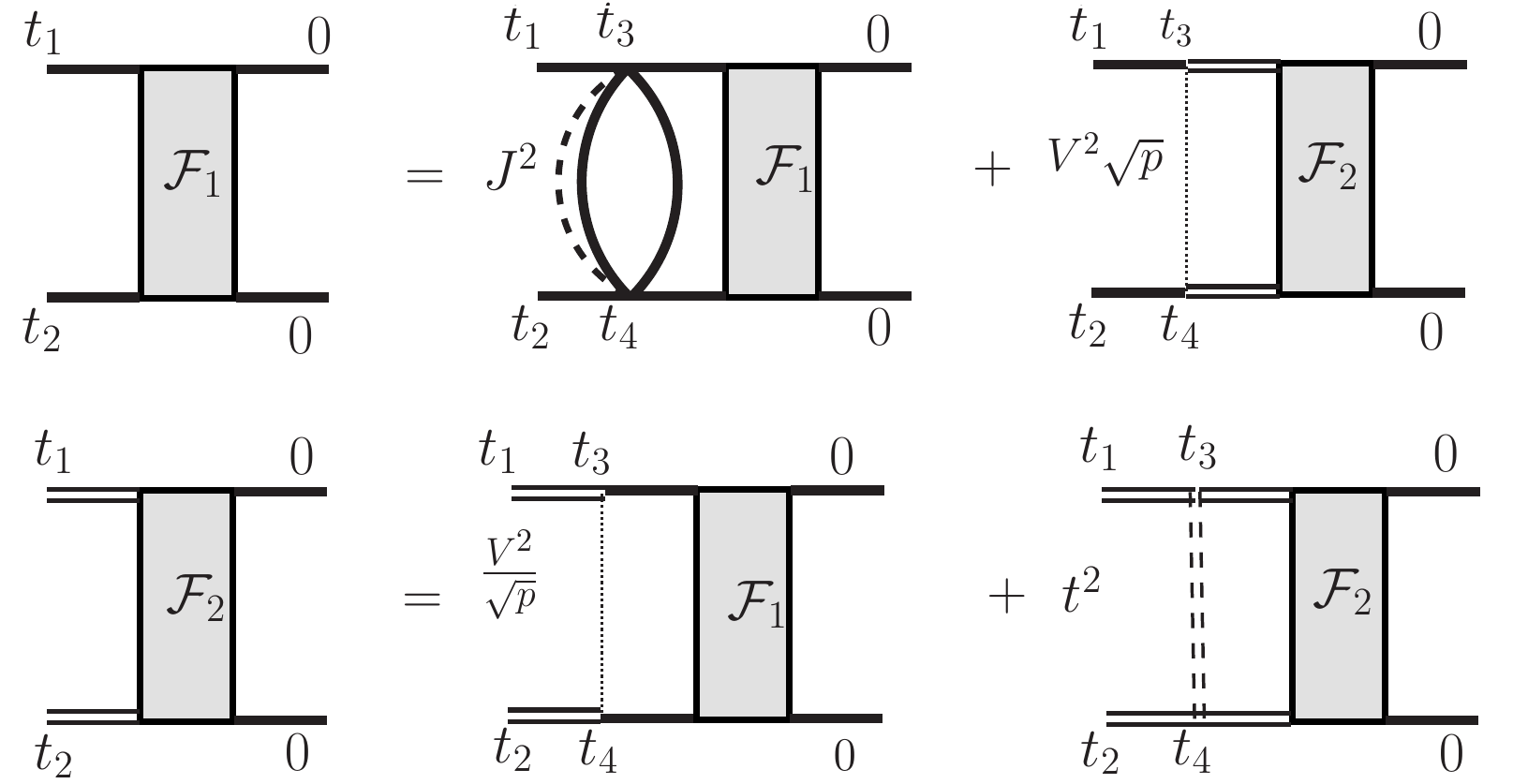}
\caption{Diagramatic illustration of the self consistency equation [Eq.\eqref{eq.BetheSalpeter}] for $1/N$ part of the OTO correlation function for Majorana fermions. Solid lines
represent the Green’s function $G$, double lines represent $\mathcal{G}$. The Kernel of Eq.\eqref{eq.Kernel} is obtained from the above diagrams for large but intermediate times $t_1,t_2$ in the chaos regime\cite{KitaevKITP}.}
\label{fig.Kernel}
\end{figure}
\end{center}

Both the four-point functions [equations \eqref{eq.OTOC}], $F=F_1,F_2$, can be obtained diagrammatically in the form (see Appendix \ref{app.OTOC})
\begin{align}
F(t_1,t_2)\simeq F^{(0)}(t_1,t_2)+\frac{1}{N}\mc{F}(t_1,t_2)+\mc{O}\left(\frac{1}{N^2}\right),
\end{align}
where $F^{(0)}$ corresponds to $\mc{O}(1)$ disconnected diagrams from contractions with the dressed propagator obtained from the saddle point equations \eqref{eq:SelfConsistency}. The $1/N$ piece $\mc{F}$ comes from ladder diagrams (see Appendix \ref{app.OTOC}). Following Refs.\onlinecite{KitaevKITP,Maldacena2016}, we obtain $\mc{F}$ via self-consistent equations represented diagrammatically in Fig. \ref{fig.Kernel},
\begin{subequations}\label{eq.BetheSalpeter}
\begin{align}
\mc{F}_1(t_1,t_2)=& \int dt_{3}dt_{4}\left[K_{11}(t_{1},t_{2},t_{3},t_{4})\mc{F}_1(t_{3},t_{4})\right.\nn\\
 &\left.+K_{12}(t_{1},t_{2},t_{3},t_{4})\mc{F}_2(t_{3},t_{4})\right]\label{eq:BetheSalpeter_1}\\
\mc{F}_2(t_1,t_2)=&  \int dt_{3}dt_{4}\left[K_{21}(t_{1},t_{2},t_{3},t_{4})\mc{F}_{1}(t_{3},t_{4})\right.\nn\\
&\left.+K_{21}(t_{1},t_{2},t_{3},t_{4})\mc{F}_{2}(t_{3},t_{4})\right]\label{eq:BetheSalpeter_2}
\end{align}
\end{subequations}
with the Kernel
\begin{align} \label{eq.Kernel}
&\mc{K}=\left(\begin{array}{cc}
K_{11} & K_{12}\\
K_{21} & K_{22}
\end{array}\right)\nn\\
\simeq&  \left(\begin{array}{cc}
3J^2G_R(t_{13})G_R(t_{24})G_{lr}^2(t_{34}) & -V^2\sqrt{p}G_R(t_{13})G_R(t_{24})\\
-\frac{V^2}{\sqrt{p}}\mc{G}_R(t_{13})\mc{G}_R(t_{24}) & -t^2\mc{G}_R(t_{13})\mc{G}_R(t_{24})
\end{array}\right)
\end{align}
Here $t_{13}=t_1-t_3$, for example, and $G_{lr}(t)\equiv iG(it+\b/2)$ is the Wightmann correlator that can be obtained by analytically continuing $G(\tau)$ via $\tau\to it+\b/2$ \cite{Maldacena2016}. 

One can recast equations\eqref{eq.BetheSalpeter} in the form of an eigenvalue equation, $\mc{K}|\mc{F}\ra=k |\mc{F}\ra$ with eigenvalue $k=1$. Anticipating chaotic dynamics we assume the following ansatz for the function $\mc{F}$
\begin{align} \label{eq.ChaosAnsatz}
|\mc{F}\ra=\left(\begin{array}{c}
\mc{F}_{1}(t_{1},t_{2})\\
\mc{F}_{2}(t_{1},t_{2})
\end{array}\right)= & e^{\l_\mrm{L}\frac{(t_{1}+t_{2})}{2}}\left(\begin{array}{c}
f_{1}(t_{12})\\
f_{2}(t_{12})
\end{array}\right).
\end{align}
The Lyapunov exponent $\l_\mrm{L}>0$ can be obtained by computing the eigenvalue $k$ and setting the condition $k(\l_\mrm{L})=1$ \cite{Maldacena2016}.

Using equation \eqref{eq.ChaosAnsatz} in equations \eqref{eq.BetheSalpeter} with the form of the Kernel in equation \eqref{eq.Kernel}, we obtain the eigenvalue equation in terms of the Fourier transforms $f_a(\w)=\int_{-\infty}^\infty dt e^{i\w t}f_a(t)$ ($a=1,2$),
\begin{subequations} \label{eq.EigenValue} 
\begin{align}
&|G_{R}\left(\tilde{\w}\right)|^{2}\left(3J^{2}\int_{-\infty}^{\infty}d\omega'g_{lr}(\omega-\omega')f_{1}(\omega')+V^{2}\sqrt{p}f_{2}(\omega)\right)\nn\\
&\hspace{13.5em}= k f_{1}(\omega)\label{eq:Eigenvaluefw1-1}\\
&|\mathcal{G}_{R}(\tilde{\w})|^{2}\left(\frac{V^{2}}{\sqrt{p}}f_{1}(\omega)+t^{2}f_{2}(\omega)\right)= k f_{2}(\omega),\label{eq:Eigenvaluefw2-1}
\end{align}
\end{subequations}
 where $g_{lr}(\w)=-\int_{-\infty}^\infty (dt/2\pi)G_{lr}^2(t)e^{i\w t}$ and $\tilde{\w}=\w+i\l_\mrm{L}/2$. 
{\em Lyapunov exponent in the NFL conformal limit --}
We now solve the above eigenvalue problem analytically in the conformal limit for $T\to 0$ in the NFL phase. Using the conformal Green's functions of equations \eqref{eq.GcFiniteT} and eliminating $f_2$ between equations \eqref{eq:Eigenvaluefw1-1} and \eqref{eq:Eigenvaluefw2-1}, we obtain a single integral equation (see Appendix \ref{app.OTOC}),
\begin{align}
&\frac{3}{4\pi}\frac{|\Gamma\left(\frac{1}{4}+\frac{h}{2}+iu\right)|^{2}}{|\Gamma\left(\frac{3}{4}+\frac{h}{2}+iu\right)|^{2}}\int_{-\infty}^{\infty}du'|\Gamma(\frac{1}{2}+i(u-u'))|^{2}f_{1}(u')\nn\\
&\hspace{10em}=\left(k-\frac{p}{k}\right)f_{1}(u),\label{eq:Kernelfw-1}
\end{align}
where $\l_\mrm{L}\equiv 2\pi hT$ and $u=\w/(2\pi T)$; $\Gamma(x)$ denotes the gamma function. As shown in Appendix \ref{app.OTOC}, the following eigenfunction solves the integral equation,
\begin{align}
f_{1}(u)= & |\Gamma\left(\frac{1}{4}+\frac{h}{2}+iu\right)|^{2},\label{eq:f1u}
\end{align}
provided that
\begin{align}
\frac{3(1-p)}{1+2h}=\left(k-\frac{p}{k}\right).\label{eq:eigenvalue}
\end{align}
The self-consistent solution of equations \eqref{eq.BetheSalpeter} is obtained by setting $k=1$ in equation \eqref{eq:eigenvalue}, leading to $h=1$, or the Lyapunov exponent
\begin{align}
\l_\mrm{L}=2\pi T
\end{align}
This implies that the entire NFL phase saturates the chaos bound for $p<1$ at half filling. This result does not apply at the QCP, since the cutoff for the conformal regime vanishes at $p=1$ (Appendix \ref{app.GcSYK}). We leave the low-temperature scrambling dynamics at the QCP for future studies.

In the time domain, equation \eqref{eq:f1u} gives 
\begin{align}
f_1(t)\propto (\cosh(\pi t/\b))^{-(h+1/2)},
\end{align}
the same as that obtained for the SYK model \cite{Maldacena2016}. The other component $f_2$, corresponding to $F_{\eta\eta\chi\chi}$ [equation \eqref{eq:OTOC2}]
\begin{align}\label{eq:f2}
f_{2}(t)\propto & \frac{JT}{V^{2}}\sqrt{\frac{p}{1-p}}\frac{1}{(\cosh(\pi t/\beta))^{(3/2+h)}},
\end{align}
is suppressed by a factor $\propto T$ relative to $f_1$ as $T\to 0$. 

\subsection{Lyapunov exponent for arbitrary filling in the NFL phase}\label{sec.OTOC.complex}
 \begin{center}
\begin{figure*}
\includegraphics[width=0.7\textwidth]{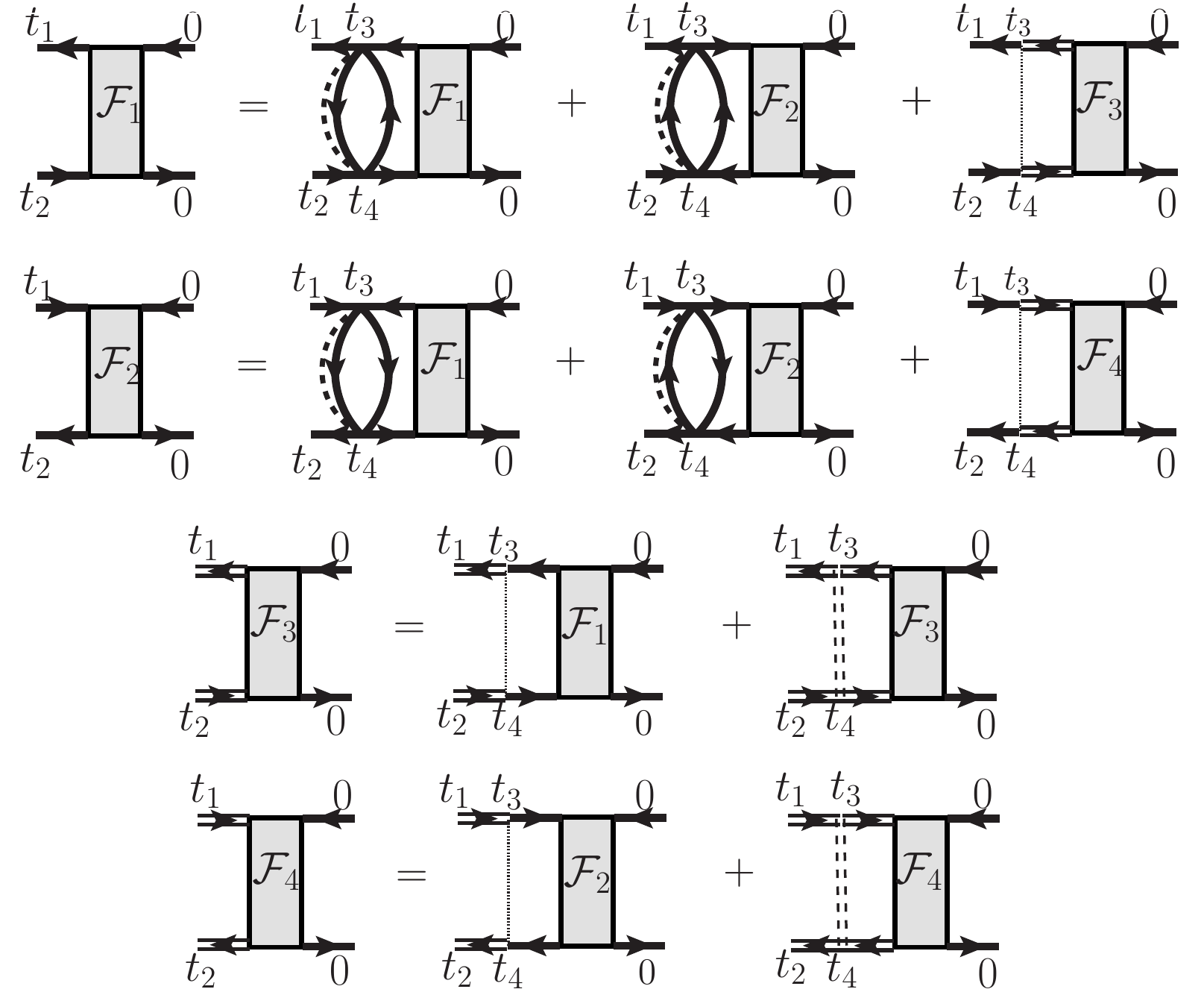}
\caption{Diagrammatic representation of the self-consistent approximation for the $1/N$ parts of the out-of-time-order functions for complex fermions.}
\label{fig.Kernel_Complex}
\end{figure*}
\end{center}  

We now generalize the calculation of the low-temperature Lyapunov exponent to arbitrary filling in the NFL phase for the complex fermion model \eqref{eq.Model}. In this case, we need to consider the following four coupled out-of-time-order
correlation functions
\begin{subequations} \label{eq:OTOC_Complex}
\begin{align}
F_{1}(t_{1},t_{2})= & \frac{1}{N^{2}}\sum_{ij}\overline{Tr[yc_{i}^{\dagger}(t_{1})yc_{j}^{\dagger}(0)yc_{i}(t_{2})yc_{j}(0)]}\\
F_{2}(t_{1},t_{2})= & \frac{1}{N^{2}}\sum_{ij}\overline{Tr[yc_{i}(t_{1})yc_{j}^{\dagger}(0)yc_{i}^{\dagger}(t_{2})yc_{j}(0)]}\\
F_{3}(t_{1},t_{2})= & \frac{1}{NM}\sum_{i\alpha}\overline{Tr[y\psi_{\alpha}^{\dagger}(t_{1})yc_{i}^{\dagger}(0)y\psi_{\alpha}(t_{2})yc_{i}(0)]}\\
F_{4}(t_{1},t_{2})= & \frac{1}{NM}\sum_{i\alpha}\overline{Tr[y\psi_{\alpha}(t_{1})yc_{i}^{\dagger}(0)y\psi_{\alpha}^{\dagger}(t_{2})yc_{i}(0)]}
\end{align}
\end{subequations}

As in the case of Majorana fermions, we estimate the $1/N$
part, $|\mc{F}\rangle=(\mc{F}_1,\mc{F}_2,\mc{F}_3,\mc{F}_4)^T$, of the above four-point functions using the self-consistent
approximation shown in Fig.\ref{fig.Kernel_Complex}. As before, this leads to the eigenvalue equation $\mc{K}|\mc{F}\rangle=k|\mc{F}\rangle$, with the self-consistent solution obtained for the eigenvalue $k=1$. The non-zero elements of the $4\times 4$ Kernel can be approximated in the chaos regime as
\begin{align*}
K_{11}= & 2J^{2}G_{A}(t_{31})G_{R}(t_{24})G_{lr}^{+}(t_{43})G_{lr}^{-}(t_{34})\\
K_{12}= & -J^{2}G_{A}(t_{31})G_{R}(t_{24})G_{lr}^{+}(t_{43})G_{lr}^{+}(t_{43})\\
K_{21}=& -J^{2}G_{R}(t_{13})G_{A}(t_{42})G_{lr}^{-}(t_{34})G_{lr}^{-}(t_{34})\\
K_{22}=& 2J^{2}G_{R}(t_{13})G_{A}(t_{42})G_{lr}^{-}(t_{34})G_{lr}^{+}(t_{43})\\
K_{13}=& V^{2}\sqrt{p}G_{A}(t_{31})G_{R}(t_{24}),~~K_{31}=\frac{V^{2}}{\sqrt{p}}\mathcal{G}_{A}(t_{31})\mathcal{G}_{R}(t_{24})\\
K_{24}=& V^{2}\sqrt{p}G_{R}(t_{13})G_{A}(t_{42}),~~K_{42}=\frac{V^{2}}{\sqrt{p}}\mathcal{G}_{R}(t_{13})\mathcal{G}_{A}(t_{42})\\
K_{33}=& t^{2}\mathcal{G}_{A}(t_{31})\mathcal{G}_{R}(t_{24}),~~K_{44}= t^{2}\mathcal{G}_{R}(t_{13})\mathcal{G}_{A}(t_{42})
\end{align*}

 There are two Wightmann correlators above, $G_{lr}^{+}(t)=iG(it+\beta/2)$
and $G_{lr}^{-}(t)=iG(it-\beta/2)$. The retarded and advanced Green's
functions are obtained as $G_{R}(t)=i\theta(t)[G(it+\eta)-G(it-\eta)]$
and $G_{A}(t)=i\theta(-t)[G(it-\eta)-G(it+\eta)]$, respectively. 

We obtain the Lyapunov exponent for $T\to 0$ by using the conformal Green's functions of Appendix \ref{app.GcSYK} in the Kernel written above. In this limit, the
elements of the Kernel are
\begin{widetext}
\begin{align}
\mathcal{K}= & \left(\begin{array}{cccc}
K_{1}e^{i\frac{\alpha}{\beta}(t_{12}-t_{34})} & \frac{1}{2}K_{1}\tan(\frac{\pi}{4}+\theta)e^{i\frac{\alpha}{\beta}(t_{12}+t_{34})} & K_{2}e^{i\frac{\alpha}{\beta}(t_{12}-t_{34})} & 0\\
\frac{1}{2}K_{1}\cot(\frac{\pi}{4}+\theta)e^{-i\frac{\alpha}{\beta}(t_{12}+t_{34})} & K_{1}e^{-i\frac{\alpha}{\beta}(t_{12}-t_{34})} & 0 & K_{2}e^{-i\frac{\alpha}{\beta}(t_{12}-t_{34})}\\
K_{3}e^{i\frac{\alpha}{\beta}(t_{12}-t_{34})} & 0 & K_{4}e^{i\frac{\alpha}{\beta}(t_{12}-t_{34})} & 0\\
0 & K_{3}e^{-i\frac{\alpha}{\beta}(t_{12}-t_{34})} & 0 & K_{4}e^{-i\frac{\alpha}{\beta}(t_{12}-t_{34})}
\end{array}\right)
\end{align}
\end{widetext}

Where $\alpha=\ln(\tan(\pi/4+\theta))$ and
\begin{subequations} 
\begin{align}
K_{1}(t_{1},t_{2},t_{3},t_{4})= & \pi(1-p)\tilde{G}_{R}(t_{13})\tilde{G}_{R}(t_{24})\tilde{G}_{lr}^{2}(t_{34})\\
K_{2}(t_{1},t_{2},t_{3},t_{4})= & \frac{V^{2}\sqrt{p}}{J}\Lambda^{2}\tilde{G}_{R}(t_{13})\tilde{G}_{R}(t_{24})\\
K_{3}(t_{1},t_{2},t_{3},t_{4})= & \frac{J\sqrt{p}\pi^{2}}{4V^{2}\Lambda^{2}}\tilde{\mathcal{G}}_{R}(t_{13})\tilde{\mathcal{G}}_{R}(t_{24})\\
K_{4}(t_{1},t_{2},t_{3},t_{4})= & \frac{t^{2}Jp\pi^{2}}{4V^{4}\Lambda^{2}}\tilde{\mathcal{G}}_{R}(t_{13})\tilde{\mathcal{G}}_{R}(t_{24})
\end{align}
\end{subequations}
 with $\tilde{G}_{R}(t)=\theta(t)/(\beta\sinh(\pi t/\beta))^{1/2}$,
$\tilde{\mathcal{G}}_{R}(t)=\theta(t)/(\beta\sinh(\pi t/\beta))^{3/2}$
and $\tilde{G}_{lr}(t)=1/(\beta\cosh(\pi t/\beta))^{1/2}$. 

We transform $\mc{K}\to\tilde{\mc{K}}$, by absorbing the exponential phase factors into eigenvector $|\mc{F}\rangle$ via the transformation $e^{i(\alpha/\beta)t_{12}}(\mc{F}_{1},\mc{F}_{3})=(\tilde{\mc{F}}_{1},\tilde{\mc{F}}_{3})$ and
$e^{-i(\alpha/\beta)t_{12}}(\mc{F}_{2},\mc{F}_{4})=(\tilde{\mc{F}}_{2},\tilde{\mc{F}}_{4})$. The eigenvalue equation $\tilde{\mathcal{K}}|\tilde{\mathcal{F}}\rangle=k|\tilde{\mathcal{F}}\rangle$
is solved by the ansatz $|\tilde{\mathcal{F}}\rangle=e^{\lambda_\mrm{L}(t_{1}+t_{2})/2}(af_{1}(t_{12}),bf_{1}(t_{12}),f_{2}(t_{12}),f_{3}(t_{12}))^T$.

 As earlier, we obtain four coupled integral equations,
\begin{subequations} \label{eq:f123}
\begin{align}
&|\tilde{G}_{R}(\tilde{\omega})|^{2}\left[\kappa_1\int_{-\infty}^{\infty}d\omega'\tilde{g}_{lr}(\omega-\omega')f_{1}(\omega')+V^{2}\sqrt{p}\Lambda^{2}f_{2}(\omega)\right]\nn\\
&\hspace{16em}= kaf_{1}(\omega) \\
&|\tilde{G}_{R}(\tilde{\omega})|^{2}\left[\kappa_2\int_{-\infty}^{\infty}d\omega'\tilde{g}_{lr}(\omega-\omega')f_{1}(\omega')+V^{2}\sqrt{p}\Lambda^{2}f_{3}(\omega)\right]\nn\\
&\hspace{16em}= kbf_{1}(\omega)\\
&\frac{\sqrt{p}\pi^{2}}{4V^{4}\Lambda^{2}}|\tilde{\mathcal{G}}_{R}(\tilde{\omega})|^{2}\left[V^{2}af_{1}(\omega)+t^{2}\sqrt{p}f_{2}(\omega)\right]= kf_{2}(\omega)\\
&\frac{\sqrt{p}\pi^{2}}{4V^{4}\Lambda^{2}}|\tilde{\mathcal{G}}_{R}(\tilde{\omega})|^{2}\left[V^{2}bf_{1}(\omega)+t^{2}\sqrt{p}f_{3}(\omega)\right]= kf_{2}(\omega)
\end{align}
\end{subequations}
 for the Fourier transforms, $f_a(\omega)$, of $f_a(t)$ ($a=1,2,3$).
Here $\tilde{\omega}=\omega+i\lambda_\mrm{L}/2$, $\kappa_1=\pi(1-p)\left(a+\frac{1}{2}\tan(\frac{\pi}{4}+\theta)b\right)$, $\kappa_2=\pi(1-p)\left(\frac{1}{2}\cot(\frac{\pi}{4}+\theta)a+b\right)$ and
\begin{subequations} 
\begin{align}
F_{R}(z)&=\int_{-\infty}^{\infty}dt F_{R}(t)e^{izt},~~~(\mathrm{Im}z>0)\nn\\
&=\frac{T^{2\Delta-1}\Gamma(\Delta-\frac{iz}{2\pi T})}{\Gamma(2\Delta)\sin(2\pi\Delta)\Gamma(1-\Delta-\frac{iz}{2\pi T})}\\
\tilde{g}_{lr}(\omega)&=\int_{-\infty}^{\infty}\frac{d\omega}{2\pi} e^{i\omega t}\tilde{G}_{lr}^{2}(\omega)=\frac{1}{2\pi^{2}}|\Gamma(\frac{1}{2}+\frac{i\omega}{2\pi T})|^{2}
\end{align}
\end{subequations}
 with $F_{R}=\tilde{G_{R}},\tilde{\mathcal{G}}_{R}$ for $\Delta=1/4$
and $\Delta=3/4$, respectively. We can eliminate $f_{2},f_{3}$ from
from the first two equations in \eqref{eq:f123} by using the last two equations. As $T\to0$, by defiing
$h=\lambda_\mrm{L}/2\pi T$ and $u=\omega/2\pi T$, we obtain
\begin{align*}
&\frac{\kappa_1}{2\pi^2}\frac{|\Gamma(\frac{1}{4}+\frac{h}{2}+iu)|^{2}}{|\Gamma(\frac{3}{4}+\frac{h}{2}+iu)|^{2}}\int_{-\infty}^{\infty}du'|\Gamma(\frac{1}{2}+i(u-u'))|^{2}f_{1}(u')\nn\\
&\hspace{12em}=(k-\frac{p}{k})af_{1}(u)\\
&\frac{\kappa_2}{2\pi^2}\frac{|\Gamma(\frac{1}{4}+\frac{h}{2}+iu)|^{2}}{|\Gamma(\frac{3}{4}+\frac{h}{2}+iu)|^{2}}\int_{-\infty}^{\infty}du'|\Gamma(\frac{1}{2}+i(u-u'))|^{2}f_{1}(u')\nn\\
&\hspace{12em}=(k-\frac{p}{k})bf_{1}(u)
\end{align*}
 The above is diagonalized in the frequency space by $f_{1}(u)=|\Gamma(\frac{1}{4}+\frac{h}{2}+iu)|^{2}$,
as in equation \eqref{eq:f1u} for Majorana fermions. As a result we obtain following eigenvalue
equation
\begin{align}
\left(\begin{array}{cc}
1 & \frac{1}{2}\tan(\frac{\pi}{4}+\theta)\\
\frac{1}{2}\cot(\frac{\pi}{4}+\theta) & 1
\end{array}\right)\left(\begin{array}{c}
a\\
b
\end{array}\right)= & l_{k}(h)\left(\begin{array}{c}
a\\
b
\end{array}\right)
\end{align}
 where $l_{k}(h)=(k-p/k)(1+2h)/(2(1-p))$. The eigenvalues
are found to be, $l_{k}(h)=1/2,3/2$. The self-consistent solution ($k=1$)
leads $h=0,1$. The latter gives rise to scrambling with the universal
Lyapunov exponent $\lambda_\mrm{L}=2\pi T$. Hence the entire NFL phase saturates the chaos bound \cite{Maldacena2015}.

\subsection{Numerical calculation of the Lyapunov exponent}\label{subsec.NumLyapunov}
 \begin{center}
\begin{figure*}
\includegraphics[width=\textwidth]{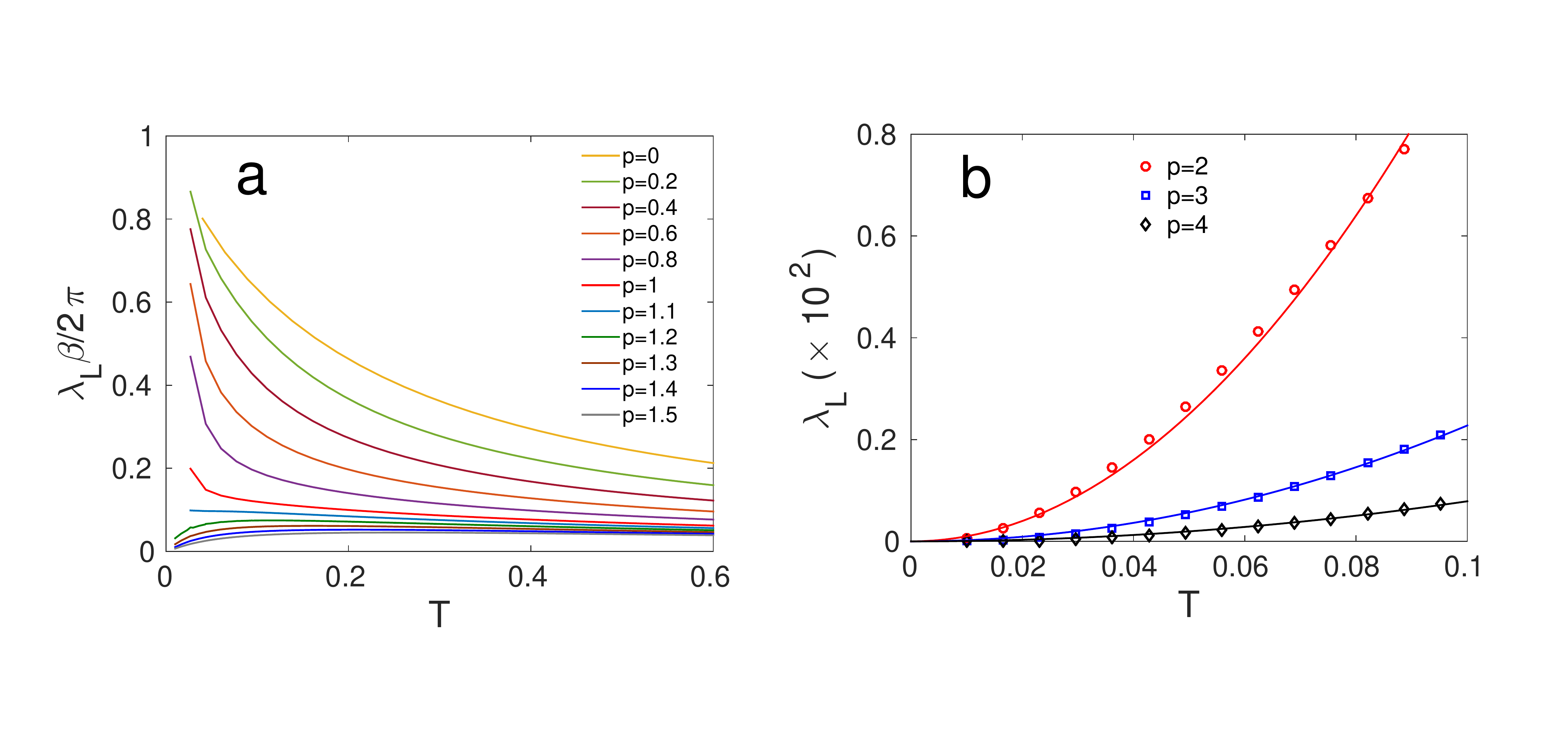}
\caption{Numerical results for the temperature dependence of the Lyupanov exponent for various values of $p$ at half filling. (a) Shows $h=\l_\mrm{L}\b/2\pi$ vs. $T$ across the QCP. The difference in the asymptotic low-temperature behaviours of $h$ in the NFL and FL phases is evident. (b) Demonstrates $T^2$ dependence of the Lyapunov exponent deep in the FL phase at low temperature. The symbols are numerical results and the lines are fits with $\l_\mrm{L}\propto T^2$.}
\label{fig.LyapunovExponent}
\end{figure*}
\end{center}
We have numerically computed $\l_\mrm{L}$ for half filling at finite temperature by solving the eigenvalue equation \eqref{eq.EigenValue} after discretization over frequency $\w$. The quantities $G_R(\w+i\l_\mrm{L}/2)$, $\mc{G}_R(\w+i\l_\mrm{L}/2)$ and $g_{lr}(\w)$ appearing in Eq.\eqref{eq.EigenValue} have been obtained from the numerical solution of the saddle-point equations \eqref{eq:SelfConsistency}. We solve for the eigenvalues $\{k(\l_\mrm{L})\}$ for a given $\l_\mrm{L}$ and look for $\l_\mrm{L}$ that satisfies $k(\l_\mrm{L})=1$. We find the eigenvalue $k=1$ to be non-degenerate.

The numerical result for $\l_\mrm{L}$ as a function of $T$ is shown in Fig.\ref{fig.LyapunovExponent}(a) over a range of $p$ across QCP. The numerical data is consistent with the ratio $h=\l_\mrm{L}\b/2\pi$ approaching 1 in the NFL phase within the temperature range that could be accessed. As shown in Fig.\ref{fig.LyapunovExponent}(b), for $p>1$, the temperature dependence of $\l_\mrm{L}$ is consistent with a $T^2$ behaviour as expected for a Fermi liquid. This is clearly evident deep in the FL phase for $p\geq 2$.

\section{Conclusions} \label{sec.Conclusion}

In this paper we introduced a solvable model that examplifies a transition between two classes of many-body quantum chaos. The model is an extension of the SYK model \cite{Sachdev1993,KitaevKITP}, with interacting fermions residing on $N$ core sites coupled to a cloud of non-interacting fermions on $M$ peripheral sites.
The model is solvable in the scaling limit $N,M\to\infty$, with $p=M/N$ kept constant. The parameter $p$ tunes the system through a quantum phase transition 
 from the non-fermi liquid state established in the pure SYK model to a fermi liquid like phase in which the peripheral fermions screen the interacting core. The residual entropy at $T\to 0$, which is non-zero in the NFL phase vanishes continuously upon crossing the critical point. The two phases represent qualitatively different classes of chaotic dynamics embodied in the structure of out-of-time-order correlations. Throughout the NFL phase the Lyapunov exponent, or scrambling rate, characterizing the emergence of chaos, saturates the quantum bound in the low temperature limit, i.e. $\lambda_\mrm{L}=2\pi T$. In the Fermi liquid phase on the other hand, the scrambling rate is perturbative in the interactions between fermions in the core giving $\lambda_\mrm{L}\propto T^2$ with a non universal pre-factor. 
 
The results we have presented  for the Lyapunov exponents on either side of the transition are invalidated at the critical point itself. Hence we postulate that this point represents a new dynamical universality class, which would be an interesting topic for further study. It would also be interesting to understand the holographic interpretation of the transition.  The non-Fermi liquid phase has an established correspondence with a quantum black hole in $AdS_2$, which plays the role of the fast scrambler. Therefore the transition to the free fixed point should correspond to a fundamental change in the geometry that eliminates the black hole. 

In the model considered here the quantum critical point separates states with fast and slow scrambling. It is natural to ask if further modification of the model is possible, that would bring the chaotic modes to a complete halt, while still allowing full analytic control in the large-$N$ limit. Such a solvable model will give us much needed insight into the nature of the many-body localization transition. 

\noindent{\em Acknowledgements --} We thank Subir Sachdev, Alexei Kitaev, Yingfei Gu, Xiao-Liang Qi, Ionut-Dragos Potirniche, Snir Gazit and Karen Micaheli for helpful discussions. This research was supported in part by the ERC synergy grant UQUAM.

\appendix
\section{Green's functions in the NFL phase}\label{app.GcSYK}
In this appendix, we obtain the zero- and finite-temperature Green's functions in the conformal limit for the SYK phase. These conformal Green's functions are used to compute the $T=0$ entropy (Section \ref{sec.LowTEntropy}) and out-of-time-ordered four-point functions (Section \ref{sec.OTOC}).

At $T=0$, the imaginary-time functions $F(\tau)=G(\tau),~\mc{G}(\tau)$ are obtained from the retarded Green's functions in equations \eqref{eq.SYKsolution} via spectral representation
\begin{subequations} \label{eq.GtauSpectral}
\begin{align}
F(\tau)= & -\int d\omega\frac{\rho(\omega)}{e^{-\beta\omega}+1}e^{-\omega\tau},\hspace{1em}(0<\tau<\beta)\nn\\
= & \int d\omega\frac{\rho(\omega)}{e^{\beta\omega}+1}e^{-\omega\tau},\hspace{1em}(-\beta<\tau<0)
\end{align}
\end{subequations}
where $\rho(\w)=\rho_c(\w),\rho_\psi(\w)$ are the spectral functions for the SYK and peripheral sites, respectively. These lead to
\begin{subequations} \label{eq.GcZeroT}
\begin{align}
G(\tau)= & -\Lambda\sin\left(\frac{\pi}{4}+\theta\right)\frac{1}{\sqrt{\pi J\tau}},\hspace{1em}0<\tau \nn\\
= & \Lambda\cos\left(\frac{\pi}{4}+\theta\right)\frac{1}{\sqrt{-\pi J\tau}},\hspace{1em}\tau<0\\
\mathcal{G}(\tau)= & -\frac{\sqrt{p}\pi J^2}{2V^{2}\Lambda}\sin\left(\frac{\pi}{4}+\theta\right)\frac{1}{(\pi J\tau)^{3/2}},\hspace{1em}0<\tau\nn\\
= & \frac{\sqrt{p}\pi J^2}{2V^{2}\Lambda}\cos\left(\frac{\pi}{4}+\theta\right)\frac{1}{(-\pi J\tau)^{3/2}},\hspace{1em}\tau<0.
\end{align}
\end{subequations}
Using equation \eqref{eq:SelfConsistency3} and the first equation above we can obtain the interaction self-energy, 
\begin{align}
\hat{\Sigma}_J(\tau)= & -\frac{J^{2}\Lambda^{3}}{2}\cos2\theta\sin\left(\frac{\pi}{4}+\theta\right)\frac{1}{(\pi J\tau)^{3/2}},\hspace{1em}0<\tau\nn\\
= & \frac{J^{2}\Lambda^{3}}{2}\cos2\theta\cos\left(\frac{\pi}{4}+\theta\right)\frac{1}{(-\pi J\tau)^{3/2}},\hspace{1em}\tau<0,
\end{align} 
which leads to equation \eqref{eq.SYKSigma}. The latter along with equations \eqref{eq.SYKG}, \eqref{eq.SYKg} constitute the self-consistent solution of equation \eqref{eq:SelfConsistency} at low energies as can be easily verified.

The conformal symmetry of equations \eqref{eq.SYKConformalSymmetry} allows to obtain the finite-$T$ Green's functions from $T=0$ results [equations \eqref{eq.GcZeroT}] via conformal transformation, e.g. $\tau=(\b/\pi)\tan(\pi\s/\b)$ \cite{Sachdev2015}, from the infinite line, $-\infty<\tau<\infty$, to the circle $0<\tau<\b$. The finite-$T$ Green's functions are
\begin{subequations} \label{eq.GcFiniteT}
\begin{align}
&G(\tau)\nn\\
&= -\Lambda \sin\left(\frac{\pi}{4}+\theta\right)\frac{g(\tau)}{\left(\beta J\sin\left(\frac{\pi\tau}{\beta}\right)\right)^{\frac{1}{2}}},~~(0<\tau<\beta)\nn\\
&= \Lambda \cos\left(\frac{\pi}{4}+\theta\right)\frac{g(\tau)}{\left(\beta J\sin\left(-\frac{\pi\tau}{\beta}\right)\right)^{\half}},~~(-\beta<\tau<0)
\end{align}
\begin{align}
&\mathcal{G}(\tau)\nn\\
&= -\frac{\sqrt{p}\pi J^2}{2V^{2}\Lambda}\sin\left(\frac{\pi}{4}+\theta\right)\frac{g(\tau)}{\left(\beta J\sin\left(\frac{\pi\tau}{\beta}\right)\right)^{\frac{3}{2}}},~~(0<\tau<\beta)\nn\\
&= \frac{\sqrt{p}\pi J^2}{2V^{2}\Lambda}\cos\left(\frac{\pi}{4}+\theta\right)\frac{g(\tau)}{\left(\beta J\sin\left(-\frac{\pi\tau}{\beta}\right)\right)^{\frac{3}{2}}},~~(-\beta<\tau<0)
\end{align}
\end{subequations}  
The factor $g(\tau)$ is related to the $U(1)$ gauge factor in equation \eqref{eq.SYKConformalSymmetry} and can be obtained by imposing the anti-periodic boundary condition $F(\tau+\b)=-F(\tau)$ on the fermionic Green's functions \cite{Sachdev2015}. This leads to $g(\tau)=e^{-\a\tau/\b}$ with $\a=\ln(\tan(\pi/4+\t))$, related to the spectral asymmetry.

\begin{center}
\begin{figure*}
\includegraphics[width=0.9\textwidth]{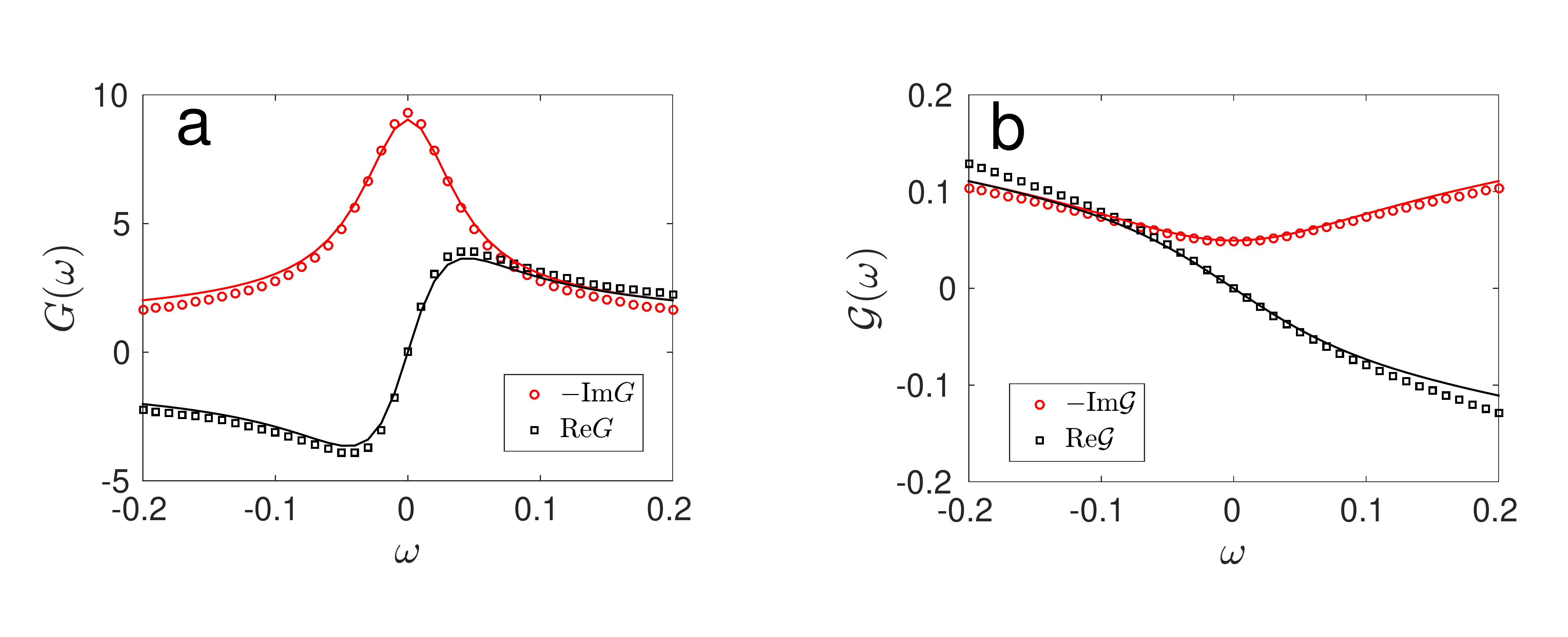}
\caption{Comparison between numerically obtained Green's functions $G$ [panel (a)] and $\mathcal{G}$ [panel (b)] at low frequency with the conformal results of equations\eqref{eq.Gzscaling} for $p=0.2$ and $T=0.025J$.}
\label{fig.GComparison}
\end{figure*}
\end{center}

The above conformal Green's functions can be rewritten in a scaling form ($0<\tau<\b$)
\begin{align} \label{eq.Gscaling}
F(\tau)=-A(\b J)^{-2\D}g\left(\frac{\tau}{\b}\right) 
\end{align}
where $g(x)=e^{-\a x}/(\sin(\pi x))^{2\D}$; $\D=\D_c,\D_\psi$, $A=A_c,A_\psi$ with $A_c=(\L/\sqrt{1+e^{-2\a}})$ and $A_\psi=(\sqrt{p}\pi J^2/(2V^2\L\sqrt{1+e^{-2\a}}))$. From these, one expects the the finite-$T$ spectral densities to also obey a scaling form
\begin{align} \label{eq.rhoscaling}
\rho(\w)=\frac{A}{J}\left(\frac{T}{J}\right)^{2\D-1}\phi\left(\frac{\w}{T}\right),
\end{align}
where $\phi$ is a scaling function. Using these scaling forms [equations \eqref{eq.Gscaling},\eqref{eq.rhoscaling}] in equation \eqref{eq.GtauSpectral}, we obtain $\phi(x)$ following the same procedure as in Ref.\onlinecite{Parcollet1998}. This gives
 \begin{align}
\phi(x)= & \frac{2^{2\Delta-1}}{\pi^{2}}e^{-\alpha/2}\cosh\left(\frac{x}{2}\right)\frac{\Gamma(\Delta+i\frac{x-\alpha}{2\pi})\Gamma(\Delta-i\frac{x-\alpha}{2\pi})}{\Gamma(2\Delta)}.
\end{align}
Here $\Gamma(x)$ is the gamma function. We obtain the conformal Green's functions on the entire complex-frequency ($z$) plane using equation \eqref{eq.rhoscaling} via the spectral representation
\begin{align} \label{eq.GzSpectral}
F(z)=\int_{-\infty}^\infty d\w\frac{\rho(\w)}{z-\w}
\end{align}
Again following Ref.\onlinecite{Parcollet1998}, using the above, we obtain $F(z)$ in a scaling form in the conformal limit for $\D<1/2$, e.g. $F(z)=G_R(z),\mc{G}_R(z)$ for $\mrm{Im}z>0$ is obtained as
\begin{subequations}\label{eq.Gzscaling}
\begin{align}
F(z)=\frac{A}{J}\left(\frac{T}{J}\right)^{2\D-1}\tilde{g}(z/T)
\end{align}
with the scaling function ($\mrm{Im}x>0$)
\begin{align} 
\tilde{g}(x)= & -i2^{2\Delta}e^{-\a/2}\frac{\cos(\pi\D+i\frac{\a}{2})}{\Gamma(2\D)\sin(2\pi\D)}\frac{\Gamma(\Delta-i\frac{x-\alpha}{2\pi})}{\Gamma(1-\Delta-i\frac{x-\alpha}{2\pi})} 
\end{align}
\end{subequations}
The integral in equation \eqref{eq.GzSpectral} has a high-frequency divergence for the conformal spectral function, $\rho_\psi$ with $\D_\psi=3/4>1/2$, for the peripheral sites. However, due to the analytical properties of gamma functions, we can analytically continue the expression in equation \eqref{eq.Gzscaling} for $\D>1/2$. As a result, equation \eqref{eq.Gzscaling} also applies for $\mc{G}_R(\w)$ in the conformal limit. We have verified that this `dimensional regularization' of the ultra-violet divergence using the fermion scaling dimension generates quite accurate $\mc{G}_R(\w)$ at low energies, when compared with the numerical results, as shown in Fig.~\ref{fig.GComparison}. We utilize this regularization in Section \ref{sec.OTOC} to calculate the four-point function in the conformal limit.

\subsection{Cutoff for the conformal solution in the NFL phase}
Here we briefly discuss the frequency cutoff for the conformal solution of equations \eqref{eq.SYKsolution} at half filling. The cutoff is estimated by comparing the terms that are neglected in the conformal limit with those that are retained in equation \eqref{eq.SYKSaddle}, at the NFL fixed point [equations \eqref{eq.SYKsolution}]. To this end, we obtain from equation \eqref{eq:SelfConsistency1} the condition that $\w\ll \w_{c1}+\w_{c2}$, where 
\begin{subequations}
\begin{align}
\omega_{c1}\simeq & \frac{J\Lambda^{6}}{2\pi^{2}}=\frac{J}{2\sqrt{\pi}}(1-p)^{3/2}\\
\omega_{c2}\simeq & \frac{p^{2}J}{2\Lambda^{2}}=\frac{J}{2\sqrt{\pi}}\frac{p^{2}}{(1-p)^{1/2}}
\end{align}
From equation \eqref{eq:SelfConsistency2}, we need to simultaneously satisfy $(V^2/\sqrt{p})G_R(\w)\gg \w,t^2\mc{G}_R(\w)$. This leads to the condition, $\w\ll \mrm{(\w_{c3},\w_{c4})}$, where
 \begin{align}
\omega_{c3}\simeq & \left(\frac{V^{4}\Lambda^{2}}{2pJ}\right)^{1/3}=\left(\frac{\sqrt{\pi}}{2}\frac{V^{4}}{J}\right)^{1/3}\left(\frac{\sqrt{1-p}}{p}\right)^{1/3}\\
\omega_{c4}\simeq & \frac{V^{4}\Lambda^{2}}{pJt^{2}}=\frac{\sqrt{\pi}V^{4}}{t^{2}J}\frac{\sqrt{1-p}}{p}
\end{align}
\end{subequations}
The overall cutoff is given by
\begin{align}
\w_c\approx\mrm{min}(\w_{c1}+\w_{c2},\w_{c3},\w_{c4})
\end{align}
For $p\to 0$, $\w_{c1}$ determines the bandwidth of the conformal behaviour, whereas $\w_{c4}$ determines the bandwidth for $p\to 1$.

\section{Luttinger theorem in the NFL phase} \label{app.LuttingerTheorem}
In this appendix we give a proof of the Luttinger theorem [equation \eqref{eq:LuttingerTheorem}] used for calculating the zero-temperature entropy in Section \ref{sec.LowTEntropy}. We perform the derivation along the same line of Refs.\onlinecite{Parcollet1998,Georges2001}.

The total fermion density $n$ at $T=0$ is given by the sum rule,
\begin{align} \label{eq.densitysumrule}
n= & -i\frac{1}{1+p}\int_{-\infty}^{\infty}\frac{d\omega}{2\pi}\left[G(\omega)+p\mathcal{G}(\omega)\right]e^{i\omega0^{+}}.
\end{align}
 $G(\omega)$ and $\mathcal{G}(\omega)$ are the time-ordered Green's
functions, e.g.~at $T=0$, $G(\omega)= \theta(\omega)G_{R}(\omega)+\theta(-\omega)G_{A}(\omega)$.
Here $A$ denotes advanced Green's function.

We rewrite equation \eqref{eq.densitysumrule}, by taking $\w$-derivative of the Dyson's equations $G^{-1}(\omega)=\omega+\mu-\Sigma(\omega)$ and $\mathcal{G}^{-1}(\omega)=\omega+\mu-\s(\omega)$, as 
\begin{align}
n =&\frac{i}{1+p}\left[\mc{P}\int_{-\infty}^{\infty}\frac{d\omega}{2\pi}\left(\frac{\partial\ln G}{\partial\omega}+p\frac{\partial\ln\mathcal{G}}{\partial\omega}\right)e^{i\omega0^{+}}\right.\nn \\
&\left.-\mc{P}\int_{-\infty}^{\infty}\frac{d\omega}{2\pi}\left(G\frac{\partial\Sigma_{G}}{\partial\omega}+p\mathcal{G}\frac{\partial\Sigma_{\mathcal{G}}}{\partial\omega}\right)e^{i\omega0^{+}}\right]\label{eq:n_1}
\end{align}
Where,
\begin{subequations}
\begin{align}
\Sigma(\omega)= & \Sigma_{J}(\omega)+V^{2}\sqrt{p}\mathcal{G}(\omega)\label{eq:Sigma_G}\\
\s(\omega)= & \frac{V^{2}}{\sqrt{p}}G(\omega)+t^{2}\mathcal{G}(\omega)\label{eq:Sigma_g}
\end{align}
\end{subequations}
 
Following procedure similar to Ref.\onlinecite{Georges2001}, we evaluate each of the terms in equation \eqref{eq:n_1} separately. From the first term in equation \eqref{eq:n_1} we get 
\begin{align}
&\mathcal{P}\int_{-\infty}^{\infty}\frac{d\omega}{2\pi}\frac{\partial\ln G}{\partial\omega}e^{i\omega0^{+}} \nn\\
&= \mathcal{P}\int_{-\infty}^{\infty}\frac{d\omega}{2\pi}\frac{\partial\ln G_{R}}{\partial\omega}e^{i\omega0^{+}}+\int_{-\infty}^{0}\frac{d\omega}{2\pi}\frac{\partial}{\partial\omega}\ln\left(\frac{G_{A}}{G_{R}}\right)e^{i\omega0^{+}}, \label{eq:integral_1}
\end{align}
and similarly for $\mathcal{G}(\omega)$. The second integral above can
be evaluated using $G_{R(A)}(\omega)=|G_{R}(\omega)|e^{\pm i\arg G_{R}(\omega)}$
leading to
\begin{align}\label{eq:integral_2}
\int_{-\infty}^{0}\frac{d\omega}{2\pi}\frac{\partial}{\partial\omega}\ln\left(\frac{G_{A}}{G_{R}}\right)e^{i\omega0^{+}}= &-i\left[\frac{1}{4}-\frac{\theta}{\pi}\right],
\end{align}
 since $\arg G_{R}(0^{-})=-(3\pi/4+\theta)$ from equation \eqref{eq.SYKsolution}
and $\arg G_{R}(-\infty)=-\pi$. Similarly
\begin{align}\label{eq:integral_3}
\int_{-\infty}^{0}\frac{d\omega}{2\pi}\frac{\partial}{\partial\omega}\ln\left(\frac{\mathcal{G}_{A}}{\mathcal{G}_{R}}\right)e^{i\omega0^{+}}= & -i\left[\frac{3}{4}+\frac{\theta}{\pi}\right],
\end{align}
using $\arg\mathcal{G}_{R}(0^{-})=-\pi/4+\theta$ from equation \eqref{eq.SYKsolution}, 

The first integral in equation \eqref{eq:integral_1} is evaluated using the conformal solution [equation \eqref{eq.SYKsolution}] as
\begin{align}\label{eq:integral_4}
\mathcal{P}\int_{-\infty}^{\infty}\frac{d\omega}{2\pi}\frac{\partial\ln G_{R}}{\partial\omega}e^{i\omega0^{+}}= &-\frac{i}{4},
\end{align}
via the deformation of the contour of integration to the upper-half plane since $G_{R}(\omega)$ is analytic there. Analogously,
\begin{align}\label{eq:integral_5}
\mathcal{P}\int_{-\infty}^{\infty}\frac{d\omega}{2\pi}\frac{\partial\ln\mathcal{G}_{R}}{\partial\omega}e^{i\omega0^{+}}= & \frac{i}{4}
\end{align}
Using equations \eqref{eq:integral_2},\eqref{eq:integral_3},\eqref{eq:integral_4},\eqref{eq:integral_5} we finally obtain
\begin{align}\label{eq:FirstTerm}
&\frac{i}{1+p}\mc{P}\int_{-\infty}^{\infty}\frac{d\omega}{2\pi}\left(\frac{\partial\ln G}{\partial\omega}+p\frac{\partial\ln\mathcal{G}}{\partial\omega}\right)e^{i\omega0^{+}}\nn \\
&= \frac{1}{1+p}\left[\left(\frac{1}{2}-\frac{\theta}{\pi}\right)+p\left(\frac{1}{2}+\frac{\theta}{\pi}\right)\right]
\end{align}

The second term in equation \eqref{eq:n_1} is rewritten using equations \eqref{eq:Sigma_G},\eqref{eq:Sigma_g} as
\begin{align}
&\mathcal{P}\int_{-\infty}^{\infty}\frac{d\omega}{2\pi}\left(G\frac{\partial\Sigma}{\partial\omega}+p\mathcal{G}\frac{\partial\sigma}{\partial\omega}\right)e^{i\omega0^{+}}\nn\\ 
&= \mathcal{P}\int_{-\infty}^{\infty}\frac{d\omega}{2\pi}\left[G\frac{\partial\Sigma_{J}}{\partial\omega}+V^{2}\sqrt{p}\frac{\partial(G\mathcal{G})}{\partial\omega}+\frac{t^{2}}{2}p\frac{\partial(\mathcal{G}^{2})}{\partial\omega}\right]
\end{align}
 The last two terms give only boundary terms that vanish. Hence, we are left with only the first term inside the bracket above. The evaluation of this term is rather cumbersome and has been done by Georges et al. \citep{Georges2001}. The calculation again only uses the information about the Green's function as $\omega\to0$ and hence does not depend on the cutoff. Using the result from Ref.\onlinecite{Georges2001} we get
\begin{align}\label{eq:SecondTerm}
i\mathcal{P}\int_{-\infty}^{\infty}\frac{d\omega}{2\pi}G\frac{\partial\Sigma_{J}}{\partial\omega} & =(1-p)\frac{\sin2\theta}{4}
\end{align}
Only difference in our case from Ref.\onlinecite{Georges2001} is the initial pre-factor
$(1-p)$, which is crucial. Finally combining the above with equation \eqref{eq:FirstTerm}, we obtain the Luttinger theorem of equation \eqref{eq:LuttingerTheorem} in Section \ref{sec.LowTEntropy}.

 \begin{center}
\begin{figure}[h!]
\includegraphics[width=0.5\textwidth]{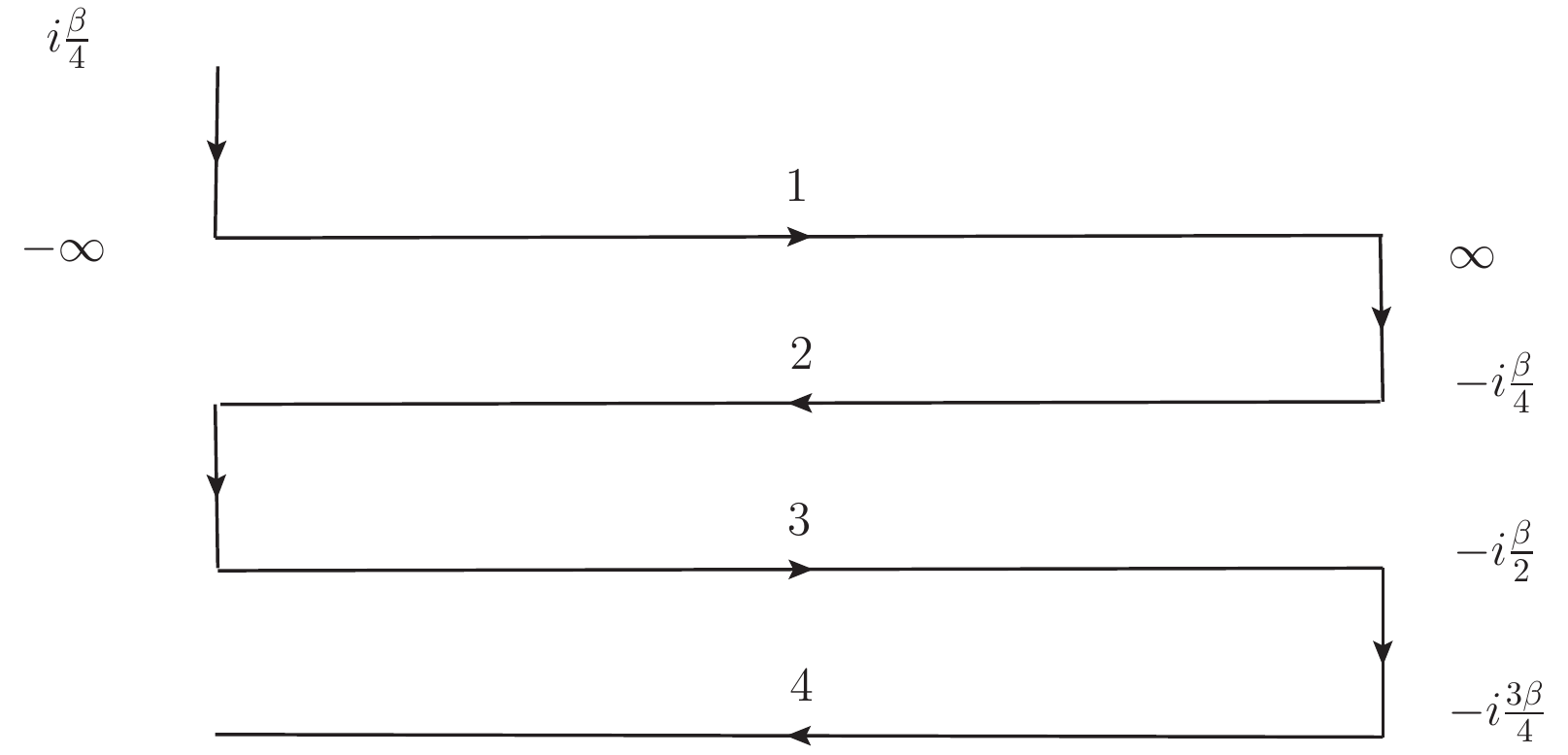}
\caption{Keldysh contour with a pair of real-time folds. Subsequent horizontal segments are separated by a quarter of the thermal cycle.}
\label{fig.KeldyshContour}
\end{figure}
\end{center} 

\begin{center}
\begin{figure*}
\includegraphics[width=0.8\textwidth]{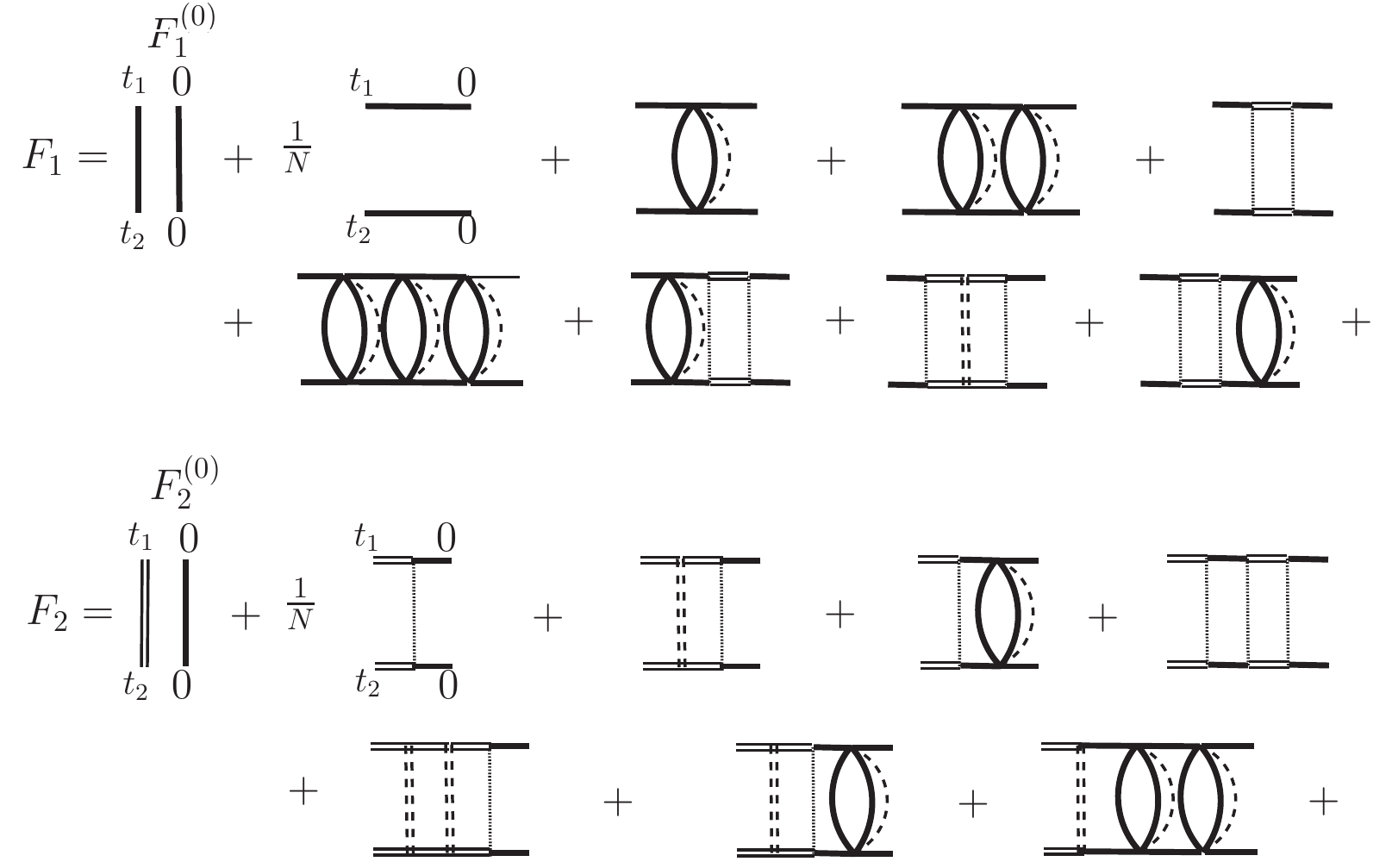}
\caption{A few lowest order diagrams corresponding to OTO correlation functions of equations\eqref{eq.OTOC}. $F_1^{(0)}$ and $F_2^{(0)}$ denote the disconnected diagrams at $\mc{O}(1)$. Only ladder diagrams contribute to the $1/N$ pieces $\mc{F}_1$ and $\mc{F}_2$, which are obtained via the self-consistent approximation of Fig.\ref{fig.Kernel}.}
\label{fig.Ladder}
\end{figure*}
\end{center}

\section{Out-of-time-ordered correlations}\label{app.OTOC}

The out-of-time-ordered correlations of equations \eqref{eq.OTOC} can be computed by formulating the problem on a Keldysh contour with four real-time segments, $\gamma=1,2,3,4$, two forward and two backward in time \cite{Stanford2016,Aleiner2016,Haehl2016}. Each of the consecutive segments are separated by quarter of the the thermal cycle \cite{Maldacena2016}. The out-of-time-order functions become contour-ordered in the Keldysh formalism, e.g. $F_1(t_1,t_2)=(1/N^2)\sum_{ij}\langle \chi_i^4(t_1)\chi_j^3(0)\chi_i^2(t_2)\chi_j^1(0)\rangle$, where $\chi^\gamma_i(t)$ denotes Grassmann variable on the branch $\gamma$. In principle, these out-of-time-order functions are coupled with four-point functions having various other time orderings, e.g. $\langle \chi_i^4(t_1)\chi_j^3(0)\chi_i^3(t_2)\chi_j^1(0)\rangle$. However, since the out-of-time-order functions decay in parametrically longer times $t\aplt (1/\lambda_\mrm{L})\ln N$ than the other four-point functions, the former effectively gets decoupled from the latter in the chaos regime.

The first few terms in diagrammatic expansions for $F_1(t_1,t_2)$ and $F_2(t_1,t_2)$ till $\mc{O}(1/N)$ are shown in Fig.\ref{fig.Ladder}. Due to disorder averaging over $\{J_{ijkl},t_{\a\b},V_{i\a}\}$, only the ladder diagrams contribute to the $1/N$ pieces $\mc{F}_1(t_1,t_2)$ and $\mc{F}_2(t_1,t_2)$. These can be obtained in the chaos regime via equations \eqref{eq.BetheSalpeter}, as discussed in Section \ref{sec.OTOC}.

The retarded functions $F_R(\w+i\l_\mrm{L}/2)$ ($F_R=G_R,\mc{G}_R$) and the Wightman correlator $G_{lr}(\w)$, appearing in equations \eqref{eq.EigenValue}, are obtained from the spectral representations
\begin{subequations}\label{eq.Kernel_Spectral}
\begin{align}
F_{R}\left(i\frac{\l_\mrm{L}}{2}+\omega\right)= & \int_{-\infty}^{\infty}d\omega'\frac{\rho(\omega')}{\omega+i\frac{\l_\mrm{L}}{2}-\omega'}\\
G_{lr}(\omega)= & -i\pi\frac{\rho_c(\omega)}{2\cosh(\beta\omega/2)}
\end{align}
 with $\rho(\omega)=\rho_c(\w),\rho_\psi(\w)$. The above leads to
\begin{align}
g_{lr}(\omega)&=-\int_{-\infty}^\infty \frac{dt}{2\pi}G_{lr}^2(t)e^{i\w t}\nn\\ &=\frac{1}{4}\int_{-\infty}^{\infty}d\omega'\frac{\rho(\omega')\rho(\omega-\omega')}{\cosh(\beta\omega'/2)\cosh(\beta(\omega-\omega')/2)}
\end{align}
\end{subequations} 
 Due to particle-hole symmetry, $G_{R}(\w+i\l_\mrm{L}/2)G_{R}(-\w+i\l_\mrm{L}/2)=-|G_{R}(\w+i\l_\mrm{L}/2)|^{2}$. This identity has been used in equation \eqref{eq.EigenValue}.

In the numerical solution of equation \eqref{eq.EigenValue}, discussed in Section \ref{sec.OTOC}, we use the spectral representations of equations \eqref{eq.Kernel_Spectral}.

\subsection{Solution of Eq.\eqref{eq.EigenValue} in the conformal limit}
In the conformal limit, we use the Green's functions of equations \eqref{eq.Gzscaling} to obtain
\begin{subequations}\label{eq.G_Kernel}
\begin{align}
G_{R}\left(i\frac{\l_\mrm{L}}{2}+i\omega\right)= &-i\frac{\L}{\sqrt{2\pi J T}}\frac{\Gamma\left(\frac{1}{4}+\frac{h}{2}-iu\right)}{\Gamma\left(\frac{3}{4}+\frac{h}{2}-iu\right)}\label{eq:GR}\\
\mathcal{G}_{R}\left(i\frac{\l_\mrm{L}}{2}+i\omega\right)= & i\frac{\sqrt{2p\pi JT}}{V^2\L}\frac{\Gamma\left(\frac{3}{4}+\frac{h}{2}-iu\right)}{\Gamma\left(\frac{1}{4}+\frac{h}{2}-iu\right)},\label{eq:gR}
\end{align}
with $\l_\mrm{L}\equiv 2\pi hT$ and $u\equiv \w/(2\pi T)$. From equation \eqref{eq.GcFiniteT}, the Wightman correlator is $G_{lr}(t)=iG(it+\b/2)=-i(\L/\sqrt{2})(\b J\cosh(\pi t/\b))^{-1/2}$ leading to
\begin{align}
g_{lr}(\w)=\frac{\L^2}{4\pi^2J}|\G(1/2+iu)|^2
\end{align}
\end{subequations}

Using equations \eqref{eq.G_Kernel} in equations \eqref{eq.EigenValue}, we get the eigenvalue equation \eqref{eq:Kernelfw-1} for $T\to 0$. The component $f_2(u)$ is obtained from $f_1(u)$,
\begin{align}
&f_2(u)=\frac{2\sqrt{\pi}JT}{V^2k}\sqrt{\frac{p}{1-p}}\frac{|\Gamma\left(\frac{3}{4}+\frac{h}{2}+iu\right)|^{2}}{|\Gamma\left(\frac{1}{4}+\frac{h}{2}+iu\right)|^{2}}f_1(u).
\end{align}
Using the form of $f_1(u)$ [equation \eqref{eq:f1u}] above we get equation \eqref{eq:f2} in Section \ref{sec.OTOC}.

To verify that $f_1(u)$ in equation \eqref{eq:f1u} is a solution of equation \eqref{eq:Kernelfw-1}, we use the identity \cite{GRBook}
\begin{align}
&\int_{-\infty}^{\infty}du'|\Gamma(\frac{1}{2}+i(u-u'))|^2|\Gamma(\frac{1}{4}+\frac{h}{2}+iu')|^2\nn\\
&=2\pi\frac{\Gamma(\frac{1}{2}+h)}{\Gamma(\frac{3}{2}+h)}|\Gamma(\frac{3}{4}+\frac{h}{2}+iu)|^{2}
\end{align}

\subsection{The functions $F_{\chi\chi\eta\eta}(t_1,t_2)$ and $F_{\eta\eta\eta\eta}(t_1,t_2)$}
We can also define two other out-of-time-ordered functions, $F_3=F_{\chi\chi\eta\eta}$ and $F_4=F_{\eta\eta\eta\eta}$, as 
\begin{subequations} \label{eq.OTOC_1}
\begin{align}
F_3(t_{1},t_{2})&= \frac{1}{NM}\sum_{i\a}\overline{\mrm{Tr}[y\chi_{i}(t_{1})y\eta_\a(0)y\chi_{i}(t_{2})y\eta_\a(0)]} \label{eq:OTOC3}\\
F_4(t_{1},t_{2})= & \frac{1}{M^2}\sum_{\alpha\b}\overline{\mrm{Tr}[y\eta_{\alpha}(t_{1})y\eta_\b(0)y\eta_{\alpha}(t_{2})y\eta_\b(0)]},
\label{eq:OTOC4}
\end{align}
\end{subequations}
 denoted by $F_3$ and $F_4$, respectively. The $1/N$ piece of these functions, $\mc{F}_3$ and $\mc{F}_4$, follow exactly same equation as in equation \eqref{eq.BetheSalpeter} with $\mc{F}_1$ replaced by $\mc{F}_3$ and $\mc{F}_2$ by $\mc{F}_4$. Hence $\mc{F}_3$ and $\mc{F}_4$ have same solutions as $\mc{F}_1$ and $\mc{F}_2$, respectively, discussed in Section \ref{sec.OTOC}. However, there is a suppression of $F_3$ by a factor $(JT/V^2)\sqrt{p/(1-p)}$ compared to $F_1$, This can be seen by evaluating the disconnected diagrams [Fig.\ref{fig.Ladder}] contributing to $F_1$ and $F_3$. 

\bibliography{SYK}

\begin{widetext}
%----------------------------------------------------------------
%----------------------------------------------------------------
\renewcommand{\thesection}{S\arabic{section}}    
\renewcommand{\thefigure}{S\arabic{figure}}
\renewcommand{\theequation}{S\arabic{equation}} 
%\makeatletter
%\renewcommand\@biblabel[1]{[S#1]}
%%\makeatletter
%%\renewcommand\citeform[1]{S#1}
%\makeatother

%\counterwithin{figure}{section}

\setcounter{figure}{0}
\setcounter{equation}{0}
\setcounter{section}{0}
%\tableofcontents
\section{Supplementary Information}
\subsection{Replica action, saddle-point equations and thermodynamic quantities}\label{app.Replica}
We obtain the saddle-point equations in equations (3) and thermodynamic quantities, such as free-energy and entropy, via the usual replica trick, $\overline{\ln Z}=\lim_{n\to 0}(\overline{Z^n}-1)/n$, where $n$ denotes number of replicas. The replicated partition function is
\begin{align}
Z^{n}= & \int\mathcal{D}(\bar{c},c)\mathcal{D}(\bar{\psi},\psi)e^{-\mathcal{S}[\bar{c},c,\bar{\psi},\psi]}.\label{eq:PartitionFn}
\end{align}
Where
\begin{subequations}\label{eq:Action}
\begin{align}
\mathcal{S=} & \mc{S}_0+\int_{0}^{\beta}d\tau\left[\frac{1}{(2N)^{3/2}}\sum_{ijkla}J_{ijkl}\bar{c}_{ia}\bar{c}_{ja}c_{ka}c_{la}+\frac{1}{M^{1/2}}\sum_{\alpha\beta a}t_{\alpha\beta}\bar{\psi}_{\alpha a}\psi_{\beta a}+\frac{1}{(NM)^{1/4}}\sum_{i\alpha a}(V_{i\alpha}\bar{c}_{ia}\psi_{\alpha a}+V_{i\alpha}^{*}\bar{\psi}_{\alpha a}c_{ia})\right],
\end{align}
with
\begin{align}
\mc{S}_0=- & \int d\tau d\tau'\left[\sum_{ia}\bar{c}_{ia}(\tau)G_{0}^{-1}(\tau,\tau')c_{ia}(\tau')+\sum_{\alpha a}\bar{\psi}_{\alpha a}(\tau)\mathcal{G}_{0}^{-1}(\tau,\tau')\psi_{\alpha a}(\tau')\right].
\end{align}
\end{subequations}
Here $a=1,\dots,n$ is the replica index and the bare Green's functions are
\begin{align}
G_{0}^{-1}(\tau,\tau')=\mathcal{G}_{0}^{-1}(\tau,\tau')= & -(\partial_{\tau}-\mu)\delta(\tau-\tau')\equiv\hat{\partial}_\tau.
\end{align}
After disorder averaging over $\{J_{ijkl},t_{\alpha\beta},V_{i\alpha}\}$ we obtain $\overline{Z^n}=\int \mc{D}(\bar{c},c)\mc{D}(\bar{\psi},\psi)e^{-\tilde{\mc{S}}[\bar{c},c,\bar{\psi},\psi]}$, where
\begin{align}
\tilde{\mc{S}}&=\mc{S}_0-\int d\tau d\tau'\sum_{ab}\left[\frac{J^2}{4N^3}\left(\sum_{ij} \bar{c}_{ia}(\tau)c_{ib}(\tau')\bar{c}_{jb}(\tau')c_{ja}(\tau)\right)^2-\frac{V^2}{\sqrt{NM}}\sum_{i\a}\bar{c}_{i a}(\tau)c_{i b}(\tau')\bar{\psi}_{\a b}(\tau')\psi_{\a a}(\tau)\right.\nn\\
&\left.-\frac{t^2}{2M}\sum_{\a\b}\bar{\psi}_{\a a}(\tau)\psi_{\a b}(\tau')\bar{\psi}_{\b b}(\tau')\psi_{\b a}(\tau)\right]
\end{align} 
We introduce the following Lagrange multiplier fields
\begin{subequations}
\begin{align}
G_{ab}(\tau,\tau')&=\frac{1}{N}\sum_i \bar{c}_{ib}(\tau')c_{ia}(\tau)\\
\mc{G}_{ab}(\tau,\tau')&=\frac{1}{M}\sum_\a \bar{\psi}_{\a b}(\tau')\psi_{\a a}(\tau)
\end{align}
\end{subequations}
by inserting of delta functions in the functional integral for $\overline{Z^n}$, i.e.
\begin{align}
\overline{Z^n}&=\int \mc{D}(\bar{c},c)\mc{D}(\bar{\psi},\psi)\mc{D}G\mc{D}\mc{G} \prod_{ab,\tau\tau'}\delta\left(NG_{ab}(\tau,\tau')-\sum_i \bar{c}_{ib}(\tau')c_{ia}(\tau)\right)\delta\left(M\mc{G}_{ab}(\tau,\tau')-\sum_\a \bar{\psi}_{\a b}(\tau')\psi_{\a a}(\tau)\right) e^{-\tilde{\mc{S}}}
\end{align}
Using the exponential representation of the delta functions via two auxiliary fields $\S_{ab}(\tau,\tau')$ and $\s_{ab}(\tau,\tau')$ and by integrating out the fermionic fields, we finally obtain
\begin{subequations}
\begin{align}
\overline{Z^n}&=\int \mc{D}[G,\mc{G},\S,\s] e^{-\mc{S}_{eff}[G,\mc{G},\S,\s]}
\end{align}
where the effective action,
\begin{align}
\mc{S}_{eff}&=-N\left[\mrm{Tr}\ln(-\hat{\partial}_\tau+\S)+ p\mrm{Tr}\ln(-\hat{\partial}_\tau+\s)\right]-N\int_0^\b d\tau d\tau'\sum_{ab}\left[\frac{J^2}{4}G_{ab}^2(\tau,\tau')G_{ba}^2(\tau',\tau)-V^2\sqrt{p}G_{ab}(\tau,\tau')\mc{G}_{ba}(\tau',\tau)\right.\nn \\
&\left.-\frac{pt^2}{2}\mc{G}_{ab}(\tau,\tau')\mc{G}_{ba}(\tau',\tau)+\S_{ab}(\tau,\tau')G_{ba}(\tau',\tau)+p \s_{ab}(\tau,\tau')\mc{G}_{ba}(\tau',\tau)\right].
\end{align}
\end{subequations}

For $N\to\infty$, the Green's functions are obtained from the saddle point of the effective action $S_{eff}$. Within replica-symmetric ansatz, where all replica-off-diagonal terms are assumed to be zero, i.e.~$X_{aa}(\tau,\tau')=X(\tau-\tau')$ and $X_{ab}(\tau,\tau')=0$ for $a\neq b$ with $X=G,\mc{G},\S,\s$, we obtain
\begin{align}
&\S(\tau)=-J^2G^2(\tau)G(-\tau)+V^2\sqrt{p}\mc{G}(\tau)\\
&\s(\tau)=\frac{V^2}{\sqrt{p}}G(\tau)+t^2\mc{G}(\tau)
\end{align}
along with the usual Schwinger-Dyson equations, $G^{-1}=\hat{\partial}_\tau-\S$ and $\mc{G}^{-1}=\hat{\partial}_\tau-\s$.
The above leads to the same self-consistency equations [equations (3)] as that obtained diagrammatically [Fig.2].

The free-energy $F=-T\overline{\ln Z}$ is evaluated at the saddle-point. This gives the free-energy density
\begin{align}
&f= \frac{F}{N+M}= -\frac{T}{1+p}\sum_{n}\left[\ln(-\beta G^{-1}(i\omega_{n}))+p\ln(-\beta\mathcal{G}^{-1}(i\omega_{n}))\right]e^{i\w_n0^+}\nonumber \\
&- \frac{1}{1+p}\int_{0}^{\beta}d\tau\left[\frac{3}{4}\Sigma_{J}(\tau)G(-\tau)+V^{2}\sqrt{p}\mathcal{G}(\tau)G(-\tau)+\frac{pt^{2}}{2}\mathcal{G}(\tau)\mathcal{G}(-\tau)\right].\label{eq:FreeEnergy}
\end{align}
 The entropy (density) is obtained as $S=-\partial f/\partial T$. To calculate entropy, we evaluate the free-energy using numerical solution of the saddle-point equations [equations (3)] for the imaginary-time Green's functions $G(\tau)$ and $\mathcal{G}(\tau)$. The Matsubara summations in equation \eqref{eq:FreeEnergy} are calculated numerically after regularizing them by addition and subtraction of the free-fermion result $-T\sum_{n}\ln(-\beta G_{0}^{-1}(i\omega_{n}))e^{i\w_n0^+}=-T\ln(1+e^{\beta\mu})$, e.g. by rewriting
\begin{align}
 &\sum_{n}\ln(-\beta G^{-1}(i\omega_{n}))e^{i\w_n0^+}=\ln(1+e^{\beta\mu})+\sum_{n}\ln\left(1-\frac{\Sigma_{J}(i\omega_{n})+V^{2}\sqrt{p}\mathcal{G}(i\omega_{n})}{i\omega_{n}+\mu}\right)
\end{align}
and similarly for the sum involving $\mc{G}^{-1}$ in equation \eqref{eq:FreeEnergy}.

\subsection{Numerical solution of the saddle point equations}\label{subsec.NumSaddlePointSoln}
To obtain the spectral functions (Section II) and the Lyapunov exponent (Section IV), we numerically solve the saddle-point equations (3) by analytical continuing to real frequency, i.e.~$i\w_n\to\w+i\eta$. This leads to the self-consistency conditions for the retarded functions,
\begin{subequations}\label{eq:SelfConsistencyR}
\begin{align}
G_R^{-1}(\omega)= & \omega+\mu-\Sigma^R_{J}(\omega)-V^{2}\sqrt{p}\mathcal{G}_R(\omega)\label{eq:SelfConsistencyR1}\\
\mathcal{G}_R^{-1}(\omega)= & \omega+\mu-\frac{V^{2}}{\sqrt{p}}G_R(\omega)-t^{2}\mathcal{G}_R(\omega)\label{eq:SelfConsistencyR2}
\end{align}
\end{subequations}
The retarded self-energy $\S_J^R(\w)$ is obtained from equation (3c) by first Fourier transforming $\Sigma_J(\tau)$ to Matsubara frequency, namely
\begin{align}
\S_J(i\w_n)&=\int_0^\b d\tau e^{i\w_n\tau}\S(\tau)=-\frac{J^2}{\b^2}\sum_{n_1,n_2}G(i\w_{n_1})G(i\w_{n_2})G(i\w_{n_1}+i\w_{n_2}-i\w_n) 
\end{align}
We carry out the above Matsubara summations using the spectral representation $G(i\w_n)=\int_{-\infty}^\infty d\w \rho_c(\w)/(i\w_n-\w)$ and analytical continue to obtain 
\begin{align} \label{eq.SelfEnergyR}
\S_J^R(\w)&=J^2\int \prod_{i=1}^3(d\w_i \rho_c(\w_i))\frac{n_\mrm{F}(-\w_1)n_\mrm{F}(\w_2)n_\mrm{F}(-\w_3)+n_\mrm{F}(\w_1)n_\mrm{F}(-\w_2)n_\mrm{F}(\w_3)}{\w-\w_1+\w_2-\w_3+i\eta}
\end{align}
The above has been used for estimating the self-energy at the FL fixed point in Section II B of the main text. To evaluate $\S_J^R(\w)$ numerically, we use the identity, $1/(\w-\w_1+\w_2-\w_3+i\eta)=-i\int_0^\infty dt e^{i(\w-\w_1+\w_2-\w_3+i\eta)t}$ to rewrite equation \eqref{eq.SelfEnergyR} as
\begin{align}
\S_J^R(\w)&=-iJ^2\int_0^\infty dt [n_1(t)n_2(t)n_1(t)+n_3(t)n_4(t)n_3(t)]e^{i\w t} \label{eq.SelfEnergy_FFT}
\end{align}
where $n_1(t)=\int_{-\infty}^\infty d\w \rho_c(\w)n_F(-\w)e^{-i\w t}$, $n_2(t)=\int_{-\infty}^\infty d\w \rho_c(\w)n_F(\w)e^{i\w t}$, $n_3(t)=\int_{-\infty}^\infty d\w \rho_c(\w)n_F(\w)e^{-i\w t}$ and $n_4(t)=\int_{-\infty}^\infty d\w \rho_c(-\w)n_F(\w)e^{-i\w t}$. 

We solve equation \eqref{eq:SelfConsistencyR} iteratively by discretizing over frequency $\w$. At each iteration step, we use fast-Fourier transform (FFT) to evaluate $\S_J^R(\w)$ and $n_a(t)$ ($a=1,\dots,4$) in equation \eqref{eq.SelfEnergy_FFT}. 

We also solve the saddle point equations (3) in imaginary time $\tau$ iteratively for the calculation of thermodynamic quantities, e.g.~the free energy [equation \eqref{eq:FreeEnergy}], discussed in subsection \ref{subsec.LowTEntropy} below. To this end, we discretize $\{G(\tau),\mc{G}(\tau)\}$ over the interval $0<\tau<\beta$. At each iteration, the Matsubara Green's functions $G(i\w_n)$, $\mathcal{G}(i\w_n)$ and the interaction self-energy $\Sigma_J(i\w_n)$ are obtained by FFT from $G(\tau),~\mc{G}(\tau)$ and $\S_J(\tau)$ [equation (3c)], respectively, by imposing anti-periodic boundary condition, $F(\tau)=-F(\tau+\beta)$, on the fermionic imaginary-time functions $F=G,\mc{G},\S_J$. After each iteration $G(i\w_n)$ is transformed back to imaginary-time via inverse-FFT in order to obtain $\S_J(\tau)$ from equation (3). 

\subsection{Quadratic fixed point away from half filling}
\begin{center}
\begin{figure*}[h!]
\includegraphics[width=\textwidth]{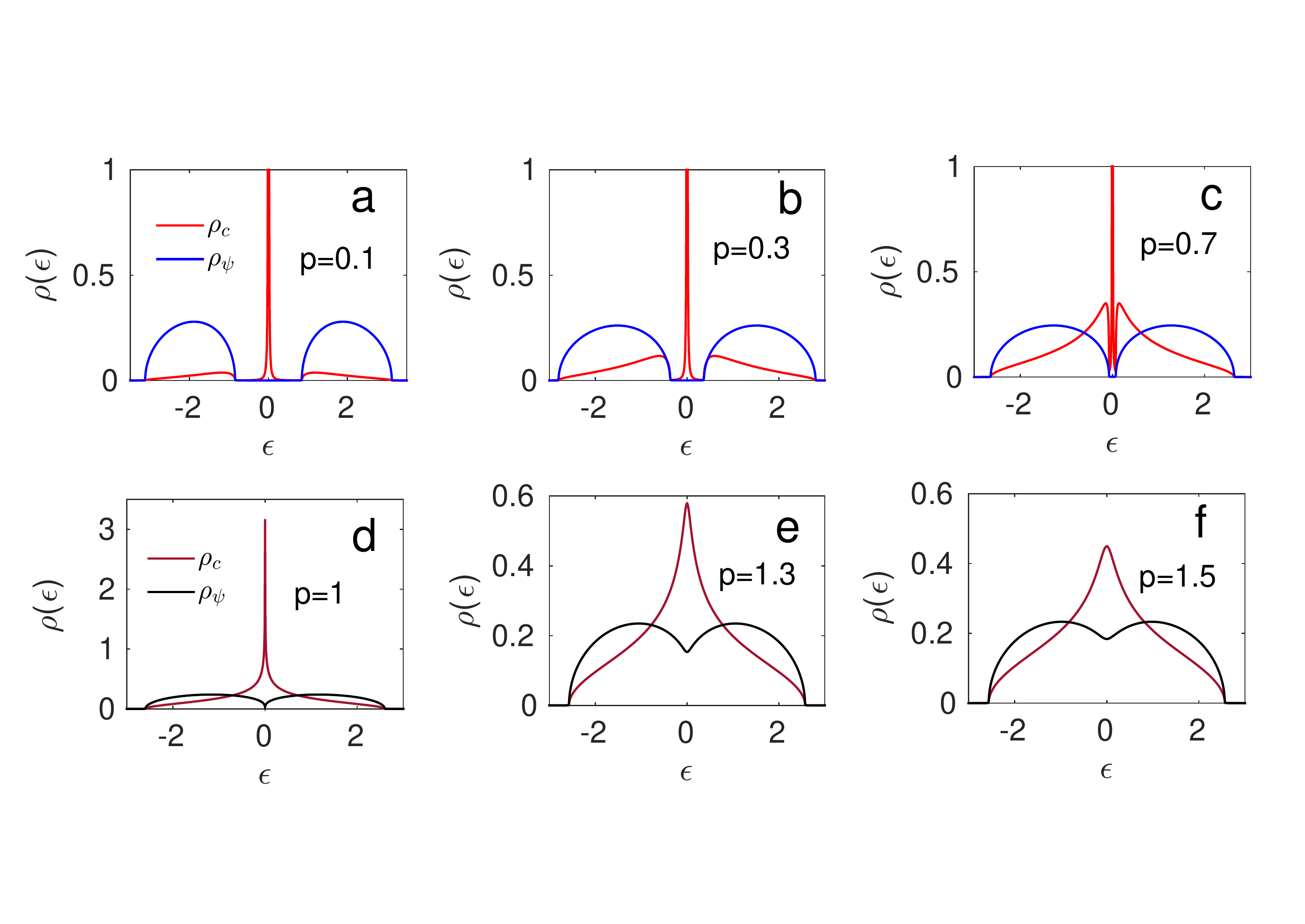}
\caption{DOS for $c$ and $\psi$ fermions as a functions of $p$ for the non-interacting case ($J=0$). For $p<1$, the spectra in panels (a)-(c) are gapped due to hybridization $V$. The gap vanishes for $p\geq 1$ as shown in panels (d)-(f).}
\label{fig.rhoNonInt}
\end{figure*}
\end{center}

In Section II B, we have shown that the model of equation (1) has a FL fixed point for $p>1$ at half filling. Here we discuss the quadratic fixed point away from half filling. To this end, we first look into the
low-energy solution in the non interacting limit by setting $J=0$ in the large-$N$ saddle-point equations (3). In this limit, the self-consistency equations are
\begin{subequations}\label{eq:NonIntSaddle} 
\begin{align}
G_{R}(\epsilon)= & \frac{1}{\e-V^{2}\sqrt{p}\mathcal{G}_{R}(\e)}\label{eq:NonintGr}\\
\mathcal{G}_{R}(\e)= & \frac{1}{\e-\frac{V^{2}}{\sqrt{p}}G_{R}(\e)-t^{2}\mathcal{G}_{R}(\e)}\label{eq:NonintgR}
\end{align}
\end{subequations}
 Here, we have defined energy, $\epsilon=\omega+\mu,$ with respect to the on site energy of the SYK sites and $\e=\mu$ corresponds to the chemical potential. For $\epsilon\to0$, we can neglect $\e$ in the equations above and obtain the constant DOS solution (10) at low energies for the FL fixed point for $p>1$, as discussed in Section II B. Similarly, it can be shown by direct substitution that following is a self-consistent solution for $\epsilon\to0$ of the above equation when $p<1$, 
\begin{subequations}
\begin{align}
G_{R}(\epsilon)\simeq & \frac{1}{\sqrt{1-p}}\frac{t}{V^{2}}-i\pi z_p\delta(\epsilon)\label{eq:GR-1}\\
\mathcal{G}_{R}^{-1}(\epsilon)\simeq & -\sqrt{\frac{p}{1-p}}t+i\pi \frac{V^2}{\sqrt{p}}z_p\delta(\epsilon)\label{eq:gR-1}
\end{align}
\end{subequations}
The delta function peak in $\rho_{c}(\e)$ at $\e=0$ corresponds to addition and removal of fermions at the SYK sites. To satisfy spectral sum rule, the delta function peak has a renormalization factor $z_p\leq 1$, that approaches one for $p\to 0$. The non-interacting Green's functions $\mc{G}_R(\e)$ of the peripherial sites vanish at $\e=0$ and both $G_{R}(\e)$ and $\mathcal{G}_{R}(\e)$ are purely real for $\epsilon\neq0$, implying that the state is an insulator for $p<1$ if $\mu$ lies close to $\epsilon=0$.

We numerically solve equations \eqref{eq:NonIntSaddle} over the entire range of $\e$.  
The resulting non-interacting DOS for $c$ and $\psi$ fermions are shown in Fig.\ref{fig.rhoNonInt}. For the sake of numerical calculation, we have added a small broadening to solve equations \eqref{eq:NonIntSaddle} and resolve the delta function peak in $\rho_c(\e)$ at $\e=0$. For $p<1$, the peak is separated by gaps from two bands, one at high and another at low energies, as shown in Figs.\ref{fig.rhoNonInt}(a),(b),(c). The spectrum of the peripheral sites is also gapped. The gaps in $\rho_c(\e)$ and $\rho_\psi(\e)$ are due to hybridization $V$. These gaps collapse to zero approaching $p=1$.

Each of the low- and high-energy bands in $\rho_c(\e)$ carries spectral weight $p/2(1+p)$ and the spectral weight under the delta-function peak is $z_p=(1-p)/(1+p)$. The spectral weight for $\rho_\psi(\e)$ is equally distributed into the two bands shown in Figs.\ref{fig.rhoNonInt}(a),(b),(c). As a result, for $p<1$, only allowed fermion densities are $n\leq p/(1+p)$, $n=1/2$ and $n\geq 1/(1+p)$, implying that the entire NFL region, except at half filling, in the $n-p$ plane of Fig.3(a) is forbidden in the absence of interaction $J$. For $n=n_c=p/(1+p),1/(1+p)$, the chemical potential lies in the gap resulting into an incompressible state for $J=0$.   

 As discussed in the main text and in the next supplementary section, turning on interaction $J$ gives rise to the NFL phase within the phase boundary $n_{c}(p)$. As demonstrated in Fig.3(b) of the main text, interaction may also destroy the incompressible state at $n_c(p)$ for $p$ close to 1, where the gaps in the non-interacting DOS become small . 

\subsection{Spectral function away from half filling}
\begin{center}
\begin{figure*}[h!]
\includegraphics[width=\textwidth]{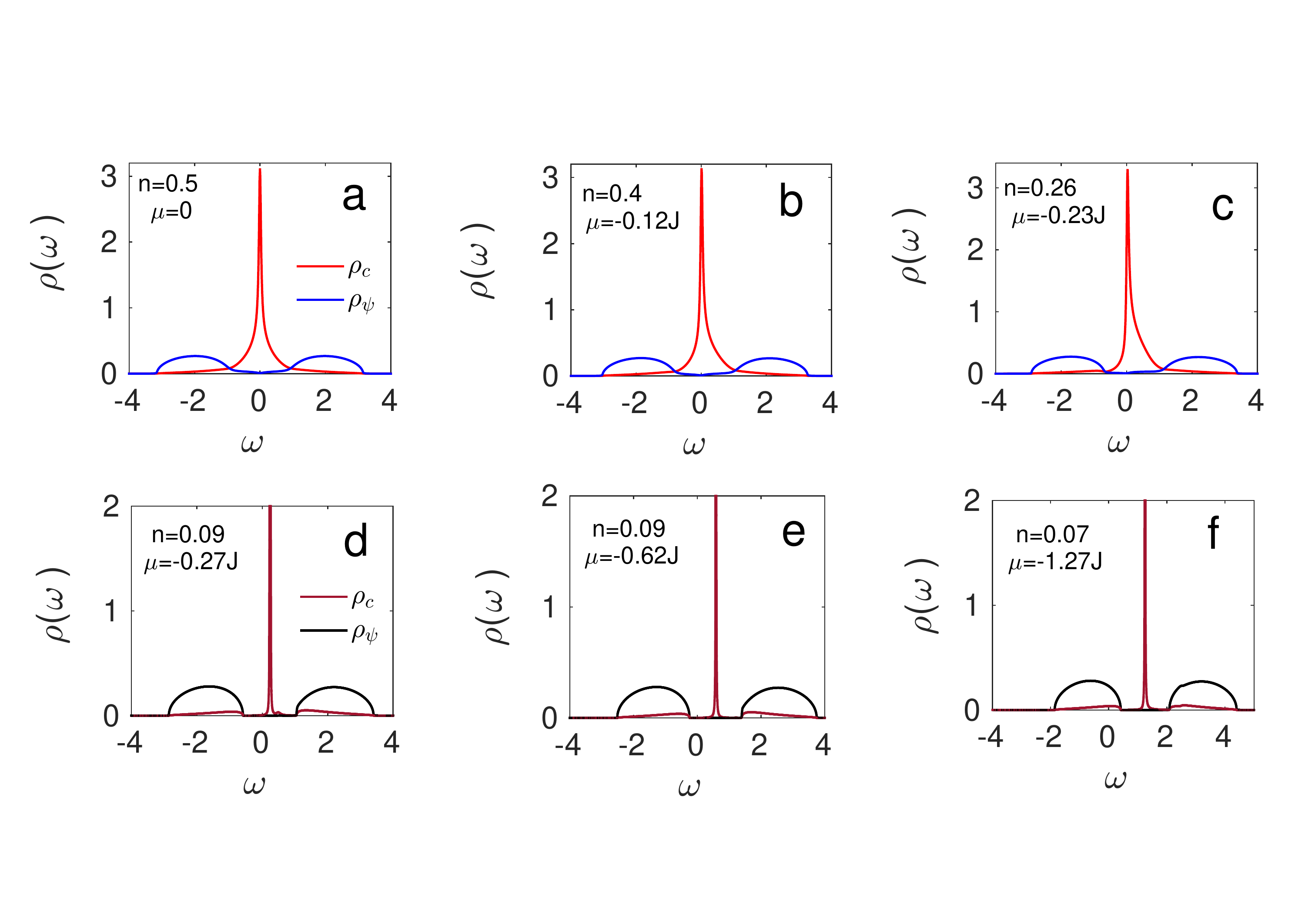}
\caption{(a)-(f) DOS of SYK fermions as a function of $\mu$ for $p=0.1$ at $T=0.025J$. The corresponding average fermion densities are indicated in the figure panels. The NFL transforms into an incompressible state at $n_c\simeq 0.09$, which corresponds to a plateau in Fig.3(b) (main text). Decreasing $\mu$ further eventaully gives rise to a metallic state.}
\label{fig.Entropy}
\end{figure*}
\end{center}

The numerical calculations of spectral functions have been done for a fixed chemical potential $\mu$, e.g.~at half filling $\mu=0$. By changing $\mu$ we can change the fermion density, $n$, and go across the phase boundary $n_c(p)$ in Fig.3 of the main text. The evolution of spectral densities, $\rho_c(\w)$ and $\rho_\psi(\w)$, across the QCP at half filling is shown in Fig.4 (main text). In Figs.\ref{fig.SpectralFunction_mu_1} and \ref{fig.SpectralFunction_mu_2}, we show the evolution of $\rho_c(\w)$ and $\rho_\psi(\w)$ as a function of $\mu$ away from half filling for two values of $p$ ($<1$). Here we only show the results for $n\leq 1/2$, i.e.~when $n$ goes through the lower phase boundary in Fig.3(a).

Depending on $J,~t$ and $V$, the NFL may transform into an incompressible state with a gap at the chemical potential, as shown for $p=0.1$ in Fig.\ref{fig.SpectralFunction_mu_1}, or a metal, as in Fig.\ref{fig.SpectralFunction_mu_2} for $p=0.7$. For both the cases, the spectral asymmetry between negative and positive frequencies in the NFL phase increases approaching $n_c$, as expected from the conformal limit (Section II A). The incompressible state is an extended region in the grand-canonical phase diagram in the $\mu-p$ plane, as indicated by the plateaus of $n$ vs.~$\mu$ in Fig.3(b) (main text). In the $n-p$ plane the incompressible state only exists at $n_c(p)$ for a range of $p$.

\begin{center}
\begin{figure*}[h!]
\includegraphics[width=\textwidth]{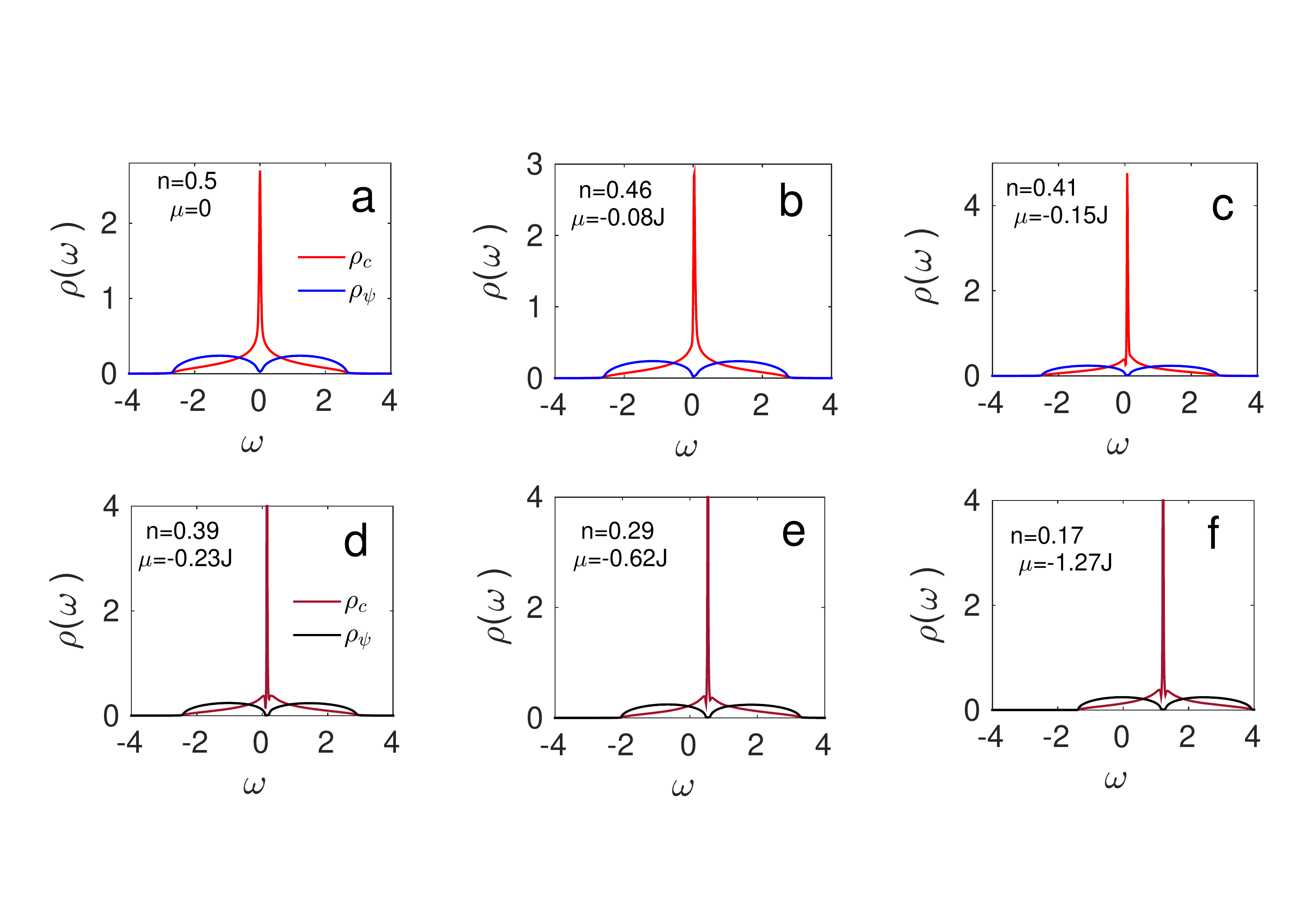}
\caption{(a)-(f) DOS of SYK fermions as a function of $\mu$ for $p=0.7$ at $T=0.025J$. The transition at $n_c\simeq 0.4$ is from NFL to a FL.}
\label{fig.SpectralFunction_mu_1}
\end{figure*}
\end{center}

\subsection{Numerical calculation of low-temperature entropy} \label{subsec.LowTEntropy}

 To verify the analytical result for zero-temperature entropy in Section III, we calculate the finite-temperature entropy, for both NFL and FL phases at half filling, using numerically obtained free-energy density $f$ [Eq.\eqref{eq:FreeEnergy}] evaluated at the saddle point. The results are  shown in Figs.\ref{fig.Entropy}. To this end, we solve the saddle-point equations (3) directly in imaginary time as discussed earlier. The numerically computation can be performed till low but finite temperature.
 
 In the SYK phase one expects the low-temperature specific heat to have $T$-linear temperature dependence due to the presence of so-called `reparameterization mode' \cite{Maldacena2016}. The latter is a quasi-Goldstone-like mode resulting from both the spontaneous breaking of conformal symmetry [equation (5)] by the SYK saddle point as well as explicit breaking of the symmetry by the non-conformal terms in equation (3) \cite{KitaevKITP,Bagrets2016,Polchinski2016,Maldacena2016}.
 
 \begin{center}
\begin{figure*}[h!]
\includegraphics[width=\textwidth]{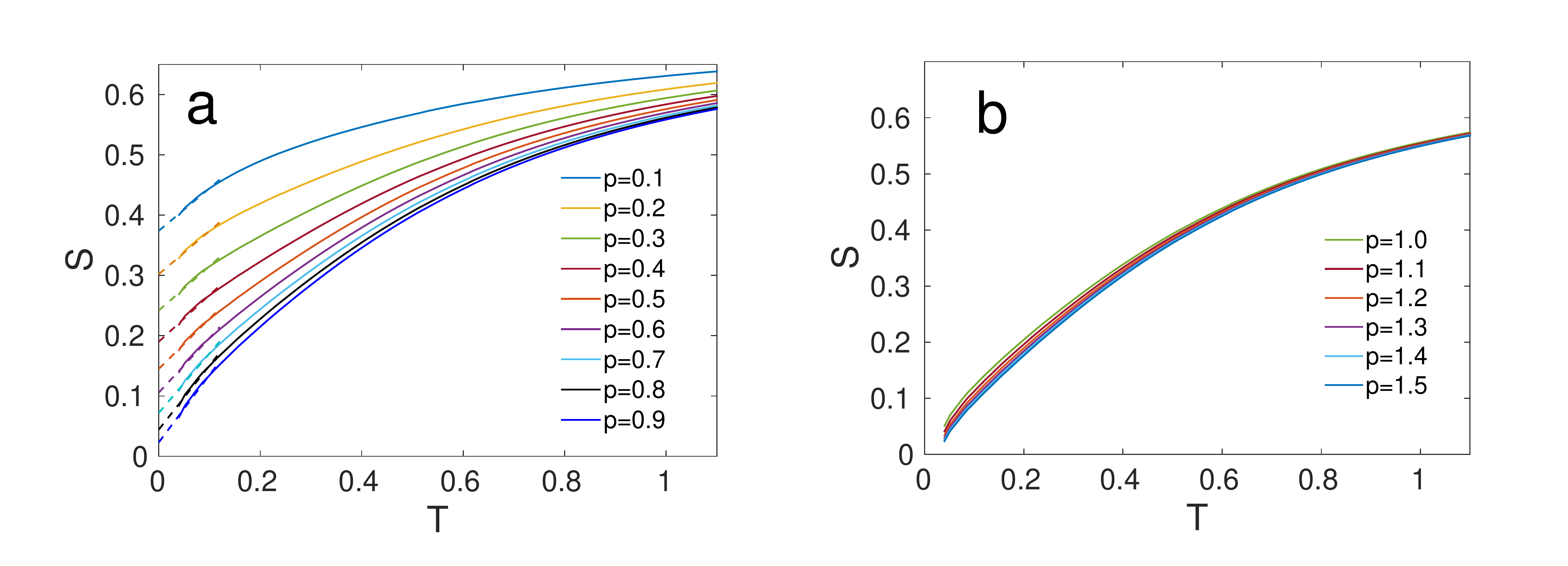}
\caption{Numerical results for finite-temperature entropy for various $p$ in the NFL [panel (a)] and FL [panel (b)] phases for $t=V=J$. The low-temperature data in the NFL phase have been linearly extrapolated to $T=0$ to estimate the zero-temperature entropy $S(T=0)$ (shown in Fig.5 of the main text).}
\label{fig.SpectralFunction_mu_2}
\end{figure*}
\end{center}
 
 The results for low-temperature entropy in the NFL phase [Fig.\ref{fig.Entropy}(a)] is consistent with a $T$-linear dependence. Hence, to estimate $T=0$ entropy, we linearly extrapolate numerically obtained entropy to $T=0$. The extrapolated value $S(T=0)$ is plotted in Fig.5 as a function of $p$ along with $S_0$ obtained analytically from equation (17). The extrapolated values for various $t=J,0.5J,0.3J,0.1J$ with $V=J$ agree quite well with the analytical results, confirming the fact that $S_0(p)$ is universal, independent of the parameters of the model, and that the $T=0$ entropy in the NFL phase vanishes continuously approaching the QCP. 
 
 We also show the results for the FL phase in Fig.\ref{fig.Entropy}(b). The low-$T$ entropy is consistent with the usual linear-$T$ behavior of FL with a zero intercept, implying $S_0=0$ in FL phase, as expected.
\end{widetext}

\end{document}